\documentstyle[epsfig,prb,aps,epic,floats]{revtex}
\newcommand{\fignorwidth}{8.6cm}
\newcommand{\fignorheight}{6cm}
\newcommand{\figwidth}{6cm}
\newcommand{\figheight}{8.6cm} 

\begin{document}

\twocolumn[\hsize\textwidth\columnwidth\hsize\csname @twocolumnfalse\endcsname

\title{
 The one-dimensional Bose-Hubbard Model with nearest-neighbor interaction
} 

\author{Till~D.~K\"uhner$^{1,2}$, Steven~R.~White$^1$ and H.~Monien$^2$} 

\address{
$^1$Department of Physics and Astronomy,University of California,Irvine, CA 92697\\
$^2$Physikalisches Institut der Universit\"at Bonn,
  D-53115 Bonn,   Germany\\
}

\date{\today}  \maketitle 

\widetext

\begin{abstract}
{
  	We study the one-dimensional Bose-Hubbard model using the 
	Density-Matrix Renormalization Group (DMRG).
	For the cases of on-site interactions and additional 
	nearest-neighbor interactions the phase boundaries
	of the Mott-insulators and charge density wave phases 
	are determined.
	We find a direct phase transition between the charge density
	wave phase and the superfluid phase, and no supersolid or 
	normal phases.
	In the presence of nearest-neighbor interaction the 
	charge density wave phase is completely surrounded by
	a region in which the effective interactions in the
	superfluid phase are repulsive. It is known from Luttinger
	liquid theory that a single impurity causes the system to be
	insulating if the effective interactions are repulsive,
	and that an even bigger region of the superfluid phase is driven 
	into a Bose-glass phase by any finite quenched disorder. 
	We determine the boundaries of both regions in the
	phase diagram. The ac-conductivity in the superfluid
	phase in the attractive and the repulsive region
	is calculated, and a big superfluid stiffness is found
	in the attractive as well as the repulsive region.
}
\end{abstract}

\vspace{8mm}

]

\narrowtext

\section{Introduction}

At zero temperature superfluid-to-insulator phase transitions can 
be found in bosonic lattice systems\cite{Fisher:1989}.
Experimental realizations of such systems are superconducting islands 
or grains connected by Josephson junctions, in which the relevant particles 
are the bosonic Cooper pairs. The transitions at zero temperature belong to the
class of quantum phase transitions that are not driven by thermal, but
by quantum fluctuations. These quantum fluctuations are controlled by
system parameters like the charging energy of the superconducting
islands and the Josephson coupling between them. Depending on such 
parameters, the system can assume different forms of long-range order.
In two dimensions superfluid-to-insulator phase transitions at zero
temperature were observed in thin granular 
films\cite{JaegerHavilandOrrGoldmann:1989,HebardPaalanen:1990,BarberDynes:1993}
and in fabricated Josephson junction arrays\cite{ZantMooij:1992,ChenDelsingHavilandHaradaCleason:1995}. 

Recently experiments were carried out in one dimensional systems
of fabricated Josephson junctions. In chains one junction wide and 63, 127 
and 255 junctions long with a tunable Josephson coupling a 
superfluid-insulator transition was 
observed\cite{Haviland:1998}. 
In another experiment with fabricated Josephson junctions, long and narrow
arrays formed an effectively one-dimensional lattice for lattice fluxes 
formed by Cooper pairs\cite{Oudenaarden:1996,Oudenaarden:1998}. The density of these 
fluxes was controlled by an external magnetic field perpendicular to 
the array, and for small ratios of Josephson coupling to charging energy, 
insulating charge density wave phases with densities $1/3$, $1/2$, $2/3$, 
$1..$ were found. 

In all of these systems the relevant particles, the Cooper pairs or the
lattice fluxes, are, at least approximately, bosonic. In this paper we will
study the phase diagram of strongly interacting bosons on one-dimensional 
lattices. In both of the above experiments on one dimensional Josephson 
junction arrays the range of the
interactions were reported to be several sites long. To study the
effect of longer-ranged interactions we will take nearest-neighbor 
interactions into account.

In the presence of on-site interactions only, Mott-insulators are found at 
integer densities, surrounded by a superfluid phase\cite{Fisher:1989}.
Additional nearest-neighbor interaction leads to charge density wave 
phases at half-integer densities. At the transition from the 
charge density wave phase to the superfluid phase two forms of 
long range order are involved: charge density wave order and 
superfluid order. If there is a direct phase transition from the
charge density wave phase to the superfluid phase, one type of order
appears at the same point where the other vanishes.
In two and higher dimensions an intermediate phase, called the
supersolid phase, that has both types of order, was found
between the charge density wave phase and the superfluid phase
in theoretical models\cite{Otterlo:1994,Batrouni:1995}.
For experimental systems the existence of supersolids is still 
controversial\cite{Meisel:1992}. In one-dimensional bosonic models 
supersolids have not been found so far\cite{Niyaz:1994}.

The existence of a normal or metallic phase that has neither 
superfluid nor charge density wave order was claimed in 
one\cite{BaltinWagenblast:1997} and two\cite{DasDoniach:1999} 
dimensional bosonic models in the high density limit.
While normal phases at zero temperature have not been ruled out rigorously, 
it has been argued that they should not exist\cite{Leggett:1973}.
We will determine whether normal or supersolid phases exist in the 
one dimensional case.

Effects of disorder and impurities can be especially strong in one-dimension. 
If the effective interactions are repulsive, a single impurity can make 
some regions of the superfluid phase insulating\cite{KaneFisher:1992}, 
while an even larger area of the
superfluid phase is turned into a lattice glass by any finite quenched 
disorder\cite{GiamarchiSchulz:1987,Fisher:1989}. 
The regions in which impurities and  disorder become relevant are 
known\cite{GiamarchiSchulz:1987,Fisher:1989,KaneFisher:1992}, 
and can be determined in the phase diagram without actually adding 
impurities or disorder to the model. Since disorder
and impurities are important in experiments, we will determine these regions
in the phase diagram. We will also compare the conductivity and Drude weight 
for the attractive and repulsive regions of the superfluid phase in pure 
systems without impurities or disorder.

The outline of this paper is as follows: in Section \ref{BoseHubbardModel}
the basic phase diagram and the possible phase transitions of the
Bose-Hubbard model are discussed. Some aspects of the density-matrix 
renormalization group are discussed in Section. \ref{DMRG}.
The calculation of the phase boundaries is presented in 
Section \ref{Phaseboundaries}.
The correlation functions in the different phases are shown in Section 
\ref{CorrelationFunctions}. In Section \ref{On-siteInteractions} 
the phase diagram with on-site interaction is presented. The possible
existence of normal or supersolid phases is discussed in 
Section \ref{NearestNeighborInteraction}, and the phase diagram
with nearest-neighbor interaction is determined. In 
Section \ref{ac-conductivity} the ac-conductivity and the superfluid 
stiffness  is calculated for the Mott-insulator and different regions 
of the superfluid phase. Conclusions are given in Section \ref{Conclusions}.

\section{The Bose-Hubbard model}
\label{BoseHubbardModel}

The basic physics of interacting bosons on a lattice is contained in
the Bose-Hubbard model\cite{Fisher:1989}.
We use an extended version which includes nearest-neighbor 
repulsion:
\begin{eqnarray}
  \label{eq.BH2}
  \nonumber
  H_{BH}  &=&  - t \sum_i ( b^{\dagger}_{i} b^{\phantom{\dagger}}_{i+1}
  +  b^{\phantom{\dagger}}_{i} b_{i+1}^{\dagger} )\\
  \nonumber
  &&  + U \sum_{i} n_{i} ( n_{i} - 1 )/2\\
  &&+ V \sum_{i}  n_{i} n_{i+1}
\end{eqnarray}
where the $b_i$ are the annihilation operators of bosons on site i,
$n_i = b_i^\dagger b_i^{\phantom{\dagger}}$ is the number of
particles on site i, and $t$ is the hopping matrix element. $U$ is the on-site 
Coulomb repulsion and $V$ is the nearest-neighbor repulsion. 
The energy scale is set by choosing $U=1$.

\begin{figure}[b]
  \begin{center}
    \epsfig{file={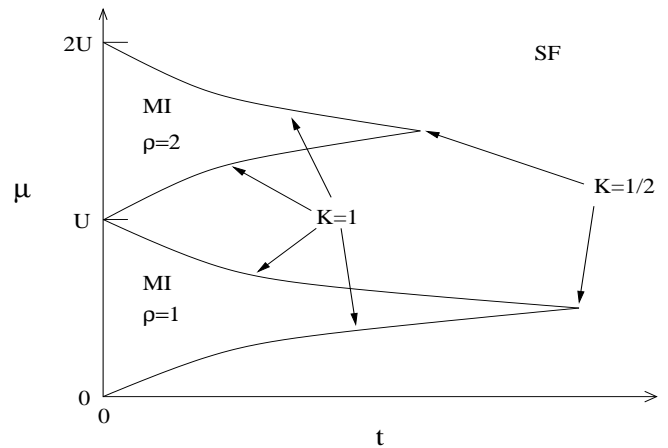},
	 height=\fignorheight,width=\fignorwidth, angle=0}
    \caption 
    {
	Schematic phase diagram with $V=0$. It shows the
	Mott-insulators (MI) with density $\rho=1$ and $\rho=2$,
	surrounded by the superfluid phase(SF). The Luttinger
	Liquid parameter $K$ is shown at the commensurate-incommensurate
	phase transitions and the Kosterlitz-Thouless transitions at
 	the tips of the insulating regions.
   }
  \label{phasediagram.ill.V=0}
  \end{center}
\end{figure}

For small interactions or large $t$ the bosons are completely delocalized,
the system is in a superfluid phase. If the density is commensurate with 
the lattice, and there is an interaction with the corresponding wavevector,
the bosons become localized at small $t$. In the presence of on-site 
interaction only ($V=0$), Mott-insulating regions with integer density 
are found. Fig. \ref{phasediagram.ill.V=0} is a sketch of the phase
diagram showing Mott-insulating regions surrounded by the superfluid
phase.

On most of the phase boundaries between the insulating phases and the
superfluid phase the density of the system changes as the phase boundary
is crossed from the incompressible insulator to the compressible
superfluid. The location of this commensurate to incommensurate density
transition can be directly determined as the energy it takes
to add a particle or hole to the insulator:
\begin{eqnarray}
  \mu_c^p   &= E^p - E_0\;,
  \label{mucplus}	\\
  - \mu_c^h &= E^h - E_0\;.	
  \label{mucminus}
\end{eqnarray}
Here $E_0$ is the energy of the insulator groundstate, $E^p$ is the
energy of a state with the density of the groundstate and an additional
particle and $E^h$ of that with an additional hole. These energies 
can be calculated using DMRG, which will be discussed in  
Section \ref{Phaseboundaries}. Note that the chemical potentials $\mu_c^p$
and $\mu_c^h$ are not equal to each other, the compressibility of the 
insulator is $\kappa = \partial \rho / \partial \mu$ is zero. 
The superfluid phase is compressible, and for states
in the superfluid phase  $\mu_c^p=\mu_c^h$. The values of $\mu_c^p$
and $\mu_c^h$ at $t=0$ shown in Fig. \ref{phasediagram.ill.V=0} can be 
easily calculated analytically.

At the phase transitions from the insulator to the superfluid phase where
the density remains an integer, the model is in the
universality class of the xy-model, and there is a Kosterlitz-Thouless\cite{Berezinskii:1972,KosterlitzThouless:1973,Kosterlitz:1974,BradleyDoniach:1984}
phase transition. This transition 
is purely driven by phase fluctuations that are determined by $t$.
The particle-hole excitation gap at the Kosterlitz-Thouless transition 
closes as:
\begin{equation}
  E_g = \mu_c^p - \mu_c^h \sim \exp{(\frac{const.}{\sqrt{t_c-t}})} \;,
\label{KT_behaviour}
\end{equation}
giving the insulating regions a very pointed shape.
The commensurate-incommensurate phase boundaries can be determined 
directly by calculating the particle and hole excitation energies 
(Eq. (\ref{mucplus}) and Eq. (\ref{mucminus})),
and in principle the Kosterlitz-Thouless transition could also
be found by locating the $t$ at which $E_g$ is zero. But since the
energy gap closes very slowly (Eq. (\ref{KT_behaviour})), small
errors in the energies lead to a big error in the location $t_c$ of the 
critical point. 
Instead, we will study the decay of the correlation functions to 
find the critical point.

The superfluid phase of interacting bosons in one dimension has a linear 
dispersion relation for small wavevectors $q$ and no excitation gap. 
The low energy physics of this
phase is that of a Luttinger Liquid\cite{Fisher:1989,Haldane:1981,Giamarchi:1992} 
with the basic Hamiltonian 
\begin{equation}
  H_0 = \frac{1}{2\pi} \int dx \left [ (v K)(\Pi(x))^2 
    + (\frac{v}{K}) (\partial_x \Phi(x))^2 \right ] \;.
\label{LuttingerHamiltonian}
\end{equation}
Here $\Pi(x)$ are density fluctuations and 
$\Phi(x)$ phase fluctuations 
($  b^\dagger(x) = \sqrt{\rho(x)}\, e^{i \Phi(x)} $), 
$v$ is the second sound velocity and $K$ determines the decay of the 
correlation function\cite{Haldane:1981}:
\begin{eqnarray}
	 \langle b^\dagger_r b_0 \rangle &&\sim r^{-K/2} 
        \label{PowerLawDecay}\\
	 \langle n_r n_0 \rangle &&\sim 1 
	+ \frac{2}{K} (2 \pi \rho r)^{-2} 
	+ A (\rho r)^{-2/K} \cos{2 \pi \rho r} \;
\label{DensityDensityFluctuationsDecay}
\end{eqnarray}
for $r\gg\rho$. The interactions and the lattice introduce an extra term:
\begin{equation}
  H_{comm} = \int dx \cos(2 n \Theta(x) + 2 \pi n x (\rho - \rho_0) ) \, ,
\label{UmklappTerm}
\end{equation}	
where $\partial_x \Theta(x) = \pi [\rho_0+ \Pi(x)]$, $\rho$ is the 
density of the system, $\rho_0$ is the density of the insulator 
(e.g. $\rho_0=1$ for the Mott insulator),
and $n$ is the denominator of the density of the insulator: $\rho_0 = m / n$.
For $K > K_c$ this term becomes relevant and drives the system into
an insulating phase.
At the Kosterlitz-Thouless transition $K_c=\frac{n^2}{2}$ and $K_c=n^2$
at the commensurate-incommensurate transition\cite{GiamarchiMillis:1992,GiamarchiPhysicaB:1997,GlazmanLarkin:1997}.  

At the Mott-insulators with integer densities, the denominator of the
density is $n=1$. At the sides of the insulator the Luttinger-Liquid
parameter is $K=1$, and at the Kosterlitz-Thouless transition at the tip 
it is $K=1/2$. The parameter $K$ can be determined from the correlation
functions, and we will locate the $t_c$ of the Kosterlitz-Thouless transition
by finding the $t$ at which $K=1/2$.

\begin{figure}[b]
  \begin{center}
    \epsfig{file={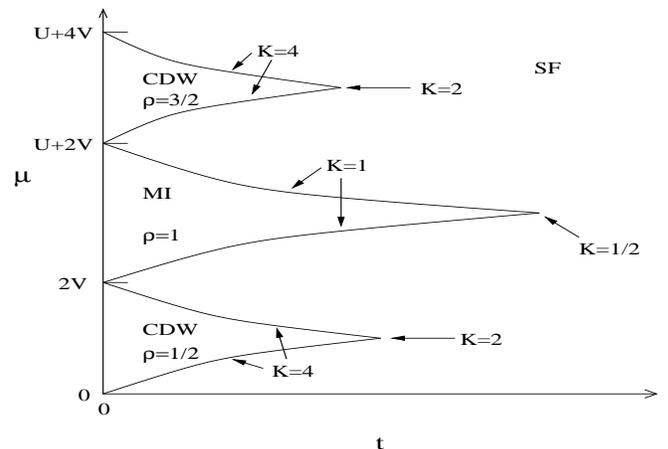},
	 height=\fignorheight,width=\fignorwidth, angle=0}
    \caption 
    {
	Schematic phase diagram with $V=0.4$. It shows the
	Mott-insulator(MI) with density $\rho=1$, the
	charge density wave (CDW) phases at densities $\rho=1/2$ and
	$\rho=3/2$, and the surrounding superfluid phase(SF). 
	The $K$ are Luttinger Liquid parameter at the phase transitions.
   }
  \label{phasediagram.ill.V=0.4}
  \end{center}
\end{figure}

If additional nearest-neighbor interactions are included in the model,
charge density wave phases are found at half integer densities. 
They also have a Kosterlitz-Thouless 
transition at the tips, giving them a similar shape as the Mott-insulators 
(Fig. \ref{phasediagram.ill.V=0.4}). Since the denominator of their density
is $n=2$, the parameter $K=4$ at the sides of the phase boundary of the 
charge density wave, and $K=2$ at the Kosterlitz-Thouless transition 
at the tip.

The possible existence of an intermediate phase, supersolid or normal, 
between the charge density wave phase and the superfluid phase will be
addressed in Section \ref{NearestNeighborInteraction}.

\section{DMRG}
\label{DMRG}

To determine the energies and correlation functions we use the Density-Matrix 
Renormalization Group (DMRG)\cite{White:1992,White:1993}, a numerical method 
capable of delivering precise results for groundstate properties of
low dimensional strongly 
interacting system. We use the finite-size version of the DMRG algorithm, 
in which the system is built up to a certain size, and the basis of the 
system is then optimized to represent the chosen target states 
by sweeping through the system repeatedly until the basis is converged.

The density matrix weight of the 
states discarded in a DMRG step is a measure of the 
numerical errors caused by the truncation. We found this truncation error 
to depend on the correlation length in the system.
At a fixed number of states kept, we find very small truncation errors in 
the insulating phases, that grow as the phase transition to the superfluid 
phase is approached, and are biggest in the superfluid phase. Note that in 
one dimension the whole superfluid phase is critical with a diverging 
correlation length, but the correlation length is always finite in finite
systems.

In each DMRG calculation for a given set of model parameters 
we first use the groundstate, the state with an additional particle
and the state with an additional hole as target states. To 
obtain adequate numerical accuracy in all cases, we require the density matrix 
weight of the truncated states $\Delta < 5 \times 10^{-6}$ (see Appendix \ref{TruncationoftheDMRGbasis}). 
The energies of these states are used to calculate the chemical potentials 
(Eq. (\ref{mucplus}) and Eq. (\ref{mucminus})).

For further sweeps only the groundstate is used as a target state. We require
the weight of the truncated states $\Delta < 10^{-9}$, and the
number of states kept is increased if necessary. After the basis is
converged, which usually takes two sweeps, the groundstate correlation
functions are calculated.

At the same parameters and number of states, the truncation error in a
system with periodic boundary conditions is usually much higher than with open
boundary conditions, therefore we use open boundary conditions. 
To keep boundary effects small we add additional terms on the boundaries to
the Hamiltonian:
\begin{equation}
  H_{boundary} = V \, \rho \, \hat{n}_1 + V \, \rho \, \hat{n}_L ;.
\end{equation}
With this additional term, a particle on the boundary on average has the same
potential energy as in the rest of the system.

\section{Phase boundaries}
\label{Phaseboundaries}

\begin{figure}[b]
  \begin{center}
    \epsfig{file={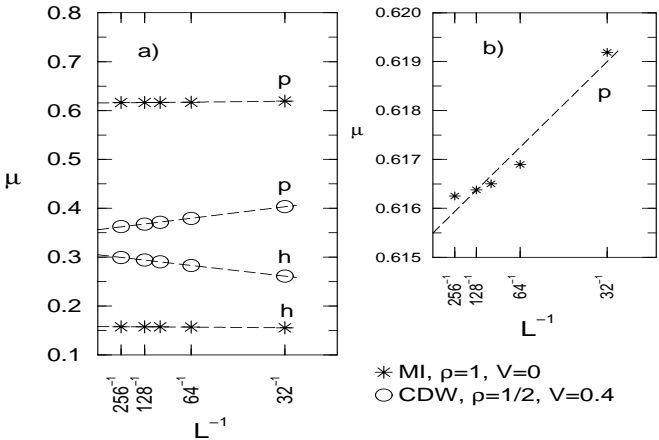},
	 height=\figheight,width=\figwidth, angle=-90}
    \caption 
    { 	a) System size dependence of the chemical potential in the 
	Mott insulator ("$*$") at $\rho=1$ with $t=0.1$ and $V=0$
	and the charge density wave phase ("$o$") at 
	$\rho=1/2$ with $t=0.1$ and $V=0.4$. 
	The upper set of data points (denoted by $p$)
	corresponds to	the energy necessary to add an additional 
	particle ($\mu_c^p$), the lower one (denoted by $h$) to that 
	of adding a hole ($\mu_c^h$). The dashed lines show linear 
	fits to the data. 
	b) $\mu^p$ for $\rho=1$, $V=0$ and $t=0.1$ on an expanded scale.
   }
  \label{ChemicalPotenial.Insulator}
  \end{center}
\end{figure}

\begin{figure}[t]
  \begin{center}
    \epsfig{file={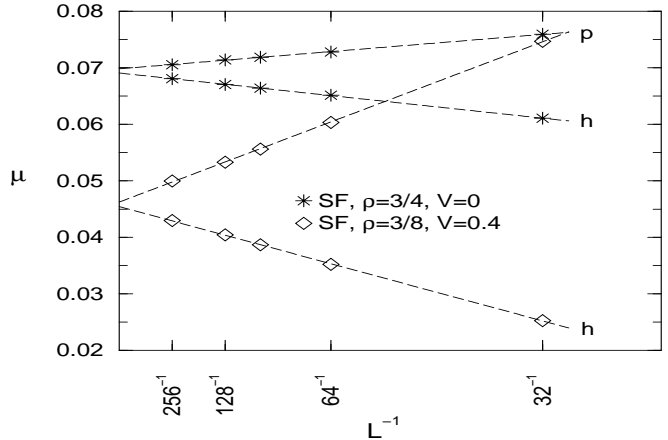},
	 height=\figheight,width=\figwidth, angle=-90}
    \caption 
    { 	System size dependence of the chemical potential in the superfluid 
     	phase at $t=0.1$. As in Fig. \ref{ChemicalPotenial.Insulator},
	the upper set of data points (p) corresponds to
	the energy necessary to add an additional particle ($\mu_c^p$), the
	lower one (h) to that of adding a hole ($\mu_c^h$). The dashed lines
 	are
	linear fits.
    }
  \label{ChemicalPotenial.Superfluid}
  \end{center}

\end{figure}

As pointed out in Section \ref{BoseHubbardModel}, the phase boundaries
with the exception of the Kosterlitz-Thouless transition 
can be determined by calculating the particle and hole excitation energies
(Eq. (\ref{mucplus}) and (\ref{mucminus})).
Using DMRG we calculate these energies in finite systems.
We expect quadratic system size dependence of the energies of the insulator
groundstates, and linear system size dependence for the states with
additional particles and holes. Fig. \ref{ChemicalPotenial.Insulator}(a) and 
Fig. \ref{ChemicalPotenial.Superfluid} show that the leading term in 
the scaling of $\mu_c^p$ and $\mu_c^h$ in the insulator and the superfluid 
phases is $1/L$. Fig. \ref{ChemicalPotenial.Insulator}(b) shows $\mu_c^p$
of the Mott-insulator at $\rho=1$ with $V=0$ and $t=0.1$. In this case
the system size dependence is very weak, and on the scale of
Fig. \ref{ChemicalPotenial.Insulator}(b) the quadratic part of the scaling
can be seen. Since the quadratic and higher parts contribute only very
weakly to the scaling, we ignore them and use linear extrapolations from 
the finite system sizes to determine $\mu_c^p$ and $\mu_c^h$ in the
thermodynamic limit. In the insulator phase  $\mu_c^p \neq \mu_c^h$, since
there is a finite gap $E_g = \mu_c^p - \mu_c^h$ (Eq. (\ref{KT_behaviour})).
In the superfluid phase the extrapolations for $\mu_c^p$ 
and $\mu_c^h$ should result in the same $\mu$ since $E_g=0$ and the system 
is compressible. In Fig. \ref{ChemicalPotenial.Superfluid} small deviations 
from this can be seen, which we will ignore in our analysis.

\section{The correlation functions}
\label{CorrelationFunctions}

\subsection{The local density}
\begin{figure}[t]
  \begin{center}
    \epsfig{file={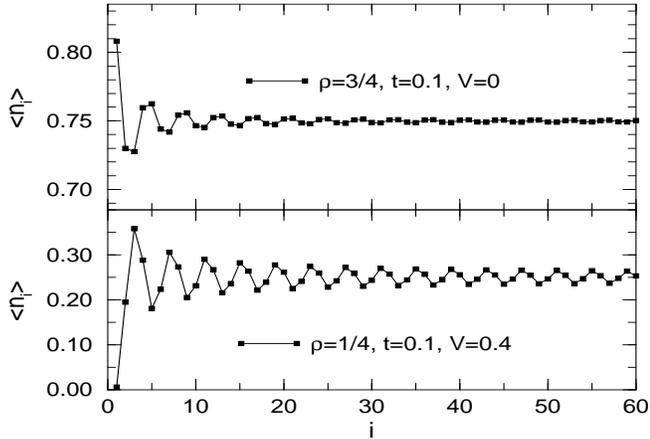},
	 height=\figheight,width=\figwidth, angle=-90}
    \caption 
    { 	Local density $\langle n_i \rangle$ in the superfluid phase.
	The systems are $L=256$ sites long.
	}
    \label{Densityfluctuations.rho0.25}		
  \end{center}
\end{figure}

Since open boundary conditions are used, special care has to be taken to
reduce boundary effects. The most obvious form of these are local density
oscillations. In the superfluid phase they show the power-law decay 
away from the edge of the system characteristic for the Luttinger Liquid. 
If the density of the bosons is given as a rational
number $\rho=n/m$, the wavelength of the oscillations is the denominator $m$ of
the density - the same wavelength as in the density-density correlation
functions (Eq. (\ref{DensityDensityFluctuationsDecay})).
Fig. \ref{Densityfluctuations.rho0.25} shows the local
density in a system at density $\rho=3/4$ with only on-site interactions,
and $\rho=1/4$ with additional nearest-neighbor interactions, both 
in the superfluid phase.

\begin{figure}[b]
  \begin{center}
    \epsfig{file={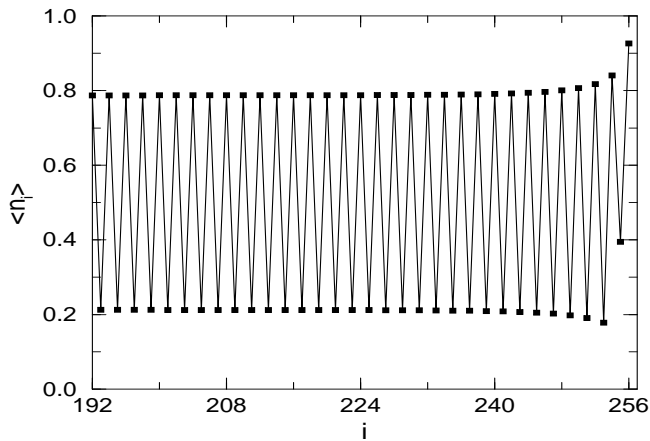},
	 height=\figheight,width=\figwidth, angle=-90}
    \caption 
    { 	Local density fluctuations in the charge density wave phase
	with density $\rho = 0.5$ at $t=0.1$, $U=1$ and $V = 0.4$
	in a $L=256$ system.
	The fluctuations induced by the right boundary at $i=256$ 
	decay quickly, but the
	true long range oscillations go throughout the system.
    }
    \label{Densityfluctuations.rho0.5}		
  \end{center}
\end{figure}

In the insulators the boundary effects decay exponentially. 
In addition to
the boundary induced density oscillations, in the charge density wave phase 
at density $\rho=1/2$ (Fig. \ref{Densityfluctuations.rho0.5}) there 
is real long range order in $\langle n_i \rangle$ in the form of a charge density wave 
with $\langle n_i \rangle = \rho + S_\pi (-1)^i$, where $S_\pi$ is the structure factor.
In an infinite system the long range order is due to
spontaneous symmetry breaking - in finite systems with even numbers of sites
we allow this by adding a small symmetry breaking term to the chemical 
potential on the left boundary.
Without this symmetry breaking term reflection symmetry would cause
the groundstate
to be the linear combination of two charge density wave phases with a
phase difference of $\pi$, canceling out the long ranged 
oscillations in $\langle n_i\rangle$. Breaking the symmetry in the finite system reduces
the Hilbert space necessary to represent the groundstate by half and 
also leads to a better convergence of the DMRG. 
In the superfluid phase the symmetry breaking term just modifies the
density oscillations at the boundary.

\subsection{The hopping correlation function}
\label{K_extrapolation}
In the superfluid phase the hopping correlation function 
$\Gamma(r) = \langle b^\dagger_r b_0 \rangle$ decays with the power-law
behavior given in Eq. (\ref{PowerLawDecay}), which can be used to  determine 
the Luttinger Liquid parameter $K$. As discussed above, there are 
local density fluctuations  $\langle n_i \rangle$ in  the finite systems. 
The creation operator can be represented by a density $\langle n_i \rangle$ 
and a phase $\phi_i$ part:
$b^\dagger_i = \sqrt{\langle n_i \rangle} \exp{(i \phi_i)}$. 
As discussed above, there are local density fluctuations
$\langle n_i \rangle$ in  the finite systems, and they will affect the 
correlation function $\langle b^\dagger_i b^{\phantom{\dagger}}_j \rangle$.
Since the local density oscillations, with the exception of the charge
density wave phase, are boundary induced, and we are interested in the
properties of an infinite system, we reduce the effect of the local density 
oscillations averaging over pairs of 
$\langle b^\dagger_i b^{\phantom{\dagger}}_j \rangle$ with 
$\mid i-j \mid = r$. To minimize boundary effects 
we place $i$ and $j$ symmetrically around the center.

\begin{figure}[tb]
  \begin{center}
    \epsfig{file={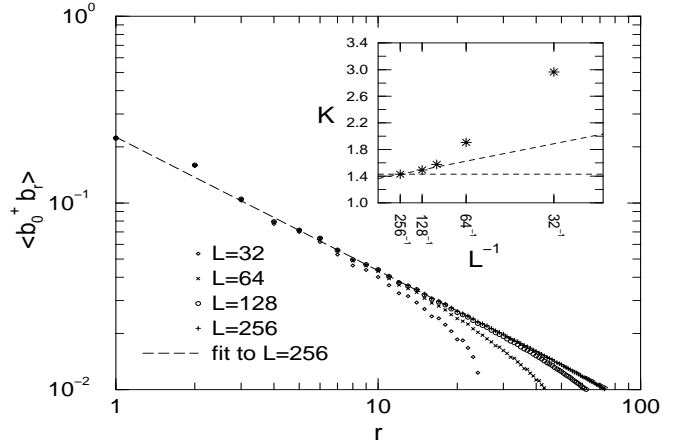},
	 height=\figheight,width=\figwidth, angle=-90}
    \caption 
    { 	$\langle b^\dagger_r b_0 \rangle$ correlation function in superfluid
	systems with density $\rho = 1/4$ at $t=0.1$, $U=1$ and $V = 0.4$
	and different system sizes $L$. 
	As the system size is increased the boundary 
	effects become weaker, and the curves for different system sizes 
	look similar over increasing regions. The inset shows the decay 
	parameter $K$ from power-law fits
	$\langle b^\dagger_r b_0 \rangle \sim r^{-K/2}$ to $16 \leq r \leq 32$
	for the different system sizes ($L=32,64,96,128$ and $256$).
	As the system size grows, $K$ converges
	to the value for infinite systems. The dashed lines show the 
	extrapolation to the upper and lower limit of $K$. The
	mean value of the upper and lower limit is taken as $K$,
	and the difference gives an estimate of the error. 
	In the shown case it is $K=1.40 \pm 0.03$.
	}
    \label{Phasecorrelationfunction.rho0.25}		
 \end{center}
\end{figure}

Fig. \ref{Phasecorrelationfunction.rho0.25} shows the power-law behavior 
in the superfluid phase for small $r$, which is modified to a faster decay 
closer to the system boundaries.
The bigger the systems are, the bigger is the region in which the 
correlation functions fit the algebraic decay, and also the region 
in which the correlation functions look
the same for different system sizes. In all cases the two biggest 
systems we calculate are at least $128$ and $256$ sites long. To estimate $K$,
we fit $a \cdot r^{-K/2}$ to the numerical data for $16 \leq r \leq 32$. 
For systems with $128$ and $256$ sites the boundary effects in this region are 
small, while the distance is also big enough to avoid short-ranged 
(non-Luttinger Liquid) effects.

With increasing system size the boundary effects get weaker, resulting in 
a decreasing $K$ that asymptotically approaches the infinite size value. 
To find a simple estimate of this we use the $K$ determined in the 
biggest system as an upper limit $K_{u}$, and the linear extrapolation
from the values in the two biggest systems as a lower limit $K_{l}$. 
We take the mean value $K=(K_{u}+K_{l})/2$ and estimate the error 
as $\Delta K = (K_{u}-K_{l})/2$.

\subsection{The density-density correlation function}
The density-density correlation functions are calculated in the same way
as the hopping correlation function. However, in this case it is
necessary to subtract the static expectation values, measuring
\mbox{$\langle n_i n_j\rangle - \langle n_i\rangle \langle n_j\rangle$}, instead of just taking \mbox{$\langle n_i n_j\rangle$}. 

\begin{figure}[b]
  \begin{center}
    \epsfig{file={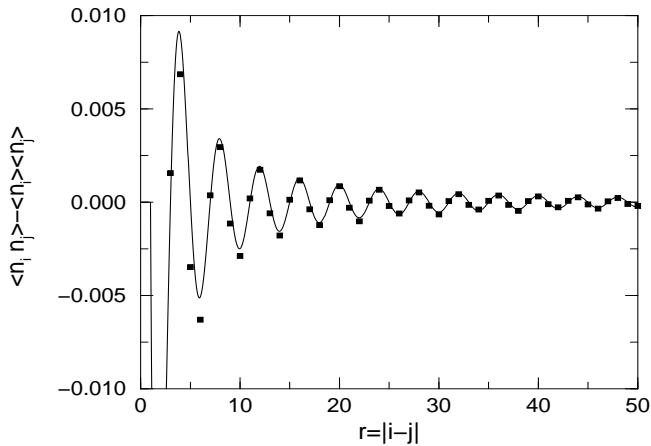},
	 height=\figheight,width=\figwidth, angle=-90}
    \caption 
    { Density-density correlation function in the superfluid phase. The boxes show the numerical data, the solid line is the fit with $A (\rho r)^{-2/K} \cos{2 \pi \rho r}$, $A=0.143$ and $K= 1.43$.
   ($\rho = 0.25$, $t=0.1$, $U=1$, $V = 0.4$, $L=256$)}
    \label{DensityDensityfluctuations.rho0.25}		
 \end{center}
\end{figure}

Fig. \ref{DensityDensityfluctuations.rho0.25} shows the density-density
correlation function at density $\rho=1/4$. A fit with Eq.
(\ref{DensityDensityFluctuationsDecay}) works fairly well, but the first
term $\frac{2}{K} (2 \pi \rho r)^{-2}$ could not be observed. 
Instead of the correlation function being bigger for small $r$, we find it to be smaller. Eq. (\ref{DensityDensityFluctuationsDecay}) only necessarily holds at large distances, and the short range behavior we see is dominated 
by the repulsive interaction between the particles. 
In fitting $A (\rho r)^{-2/K} \cos{2 \pi \rho r}$ to the data, 
a cut-off at small distances has to be made. While this works well enough 
to confirm that the correlation functions decay with a power-law behavior, 
the uncertainties in the fit are too high to determine $K$.

\section{On-site interactions}
\label{On-siteInteractions}

\begin{figure}[b]
  \begin{center}
    \protect\epsfig{file=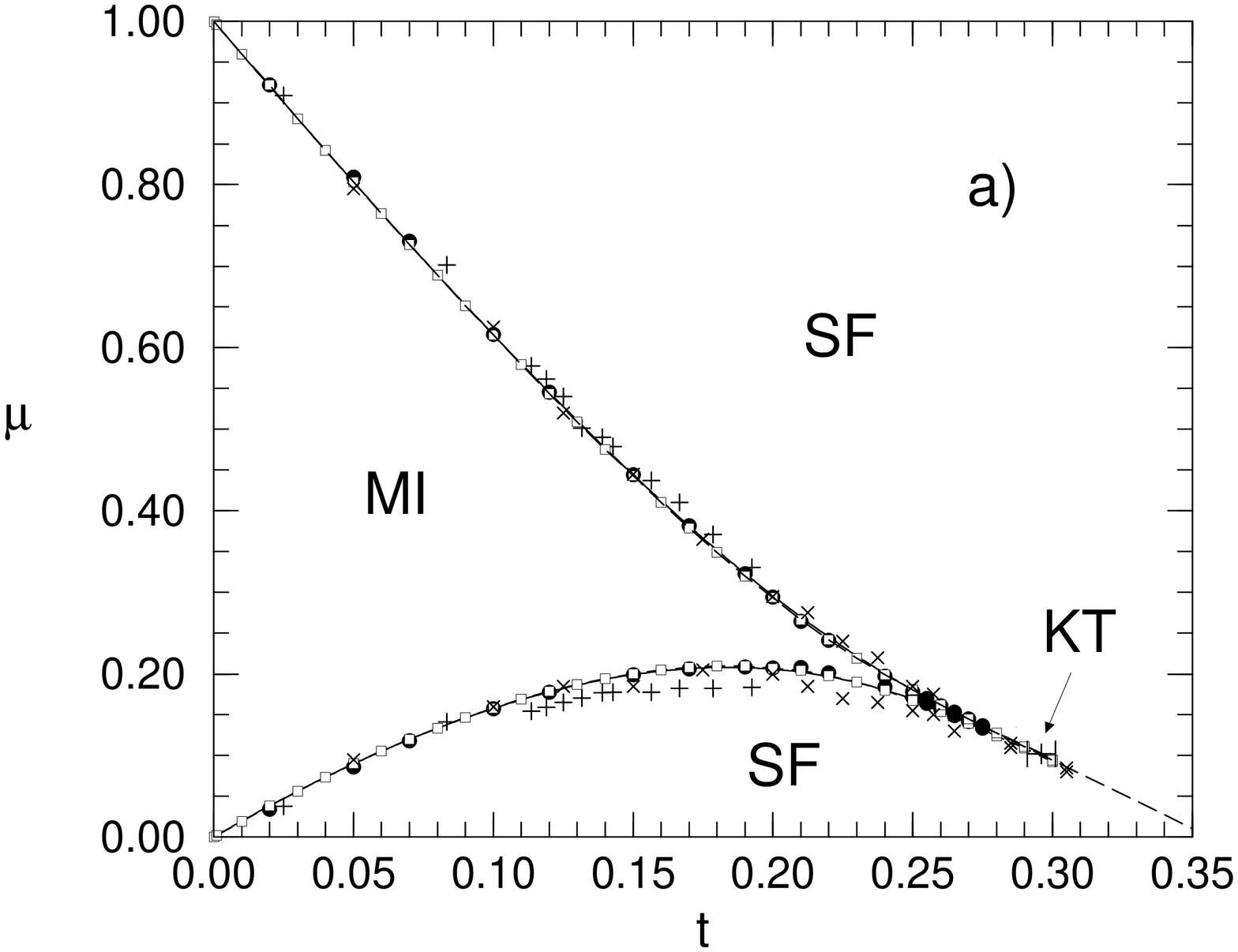,
      height=\figheight,width=\figwidth,angle=-90}
   \protect\caption[]{ The phase
      diagram with on-site interactions only (MI:Mott-insulator with
      density one, SF:superfluid phase).  The solid lines
      show a Pad\'e analysis of 
      12th order strong coupling expansions \cite{ElstnerMonien:1999},
      two different sets of Quantum Monte Carlo data are 
      ``+''\cite{Batrouni:1992} and ``x'' \cite{Kashurnikov:1996.2}.
      The filled circles show older DMRG results\cite{KuehnerMonien:1998}, 
	the empty boxes are the new DMRG data. The dashed lines indicate the
      area with integer density. The error bars in the $\mu$ direction are 
        smaller than the circles, the error bar in the $t$ direction
         is the error of $t_c$ for the Kosterlitz-Thouless (KT) transition.
	} 
	\label{Fig.phasediagram.MI.noU1}
    \protect\epsfig{file=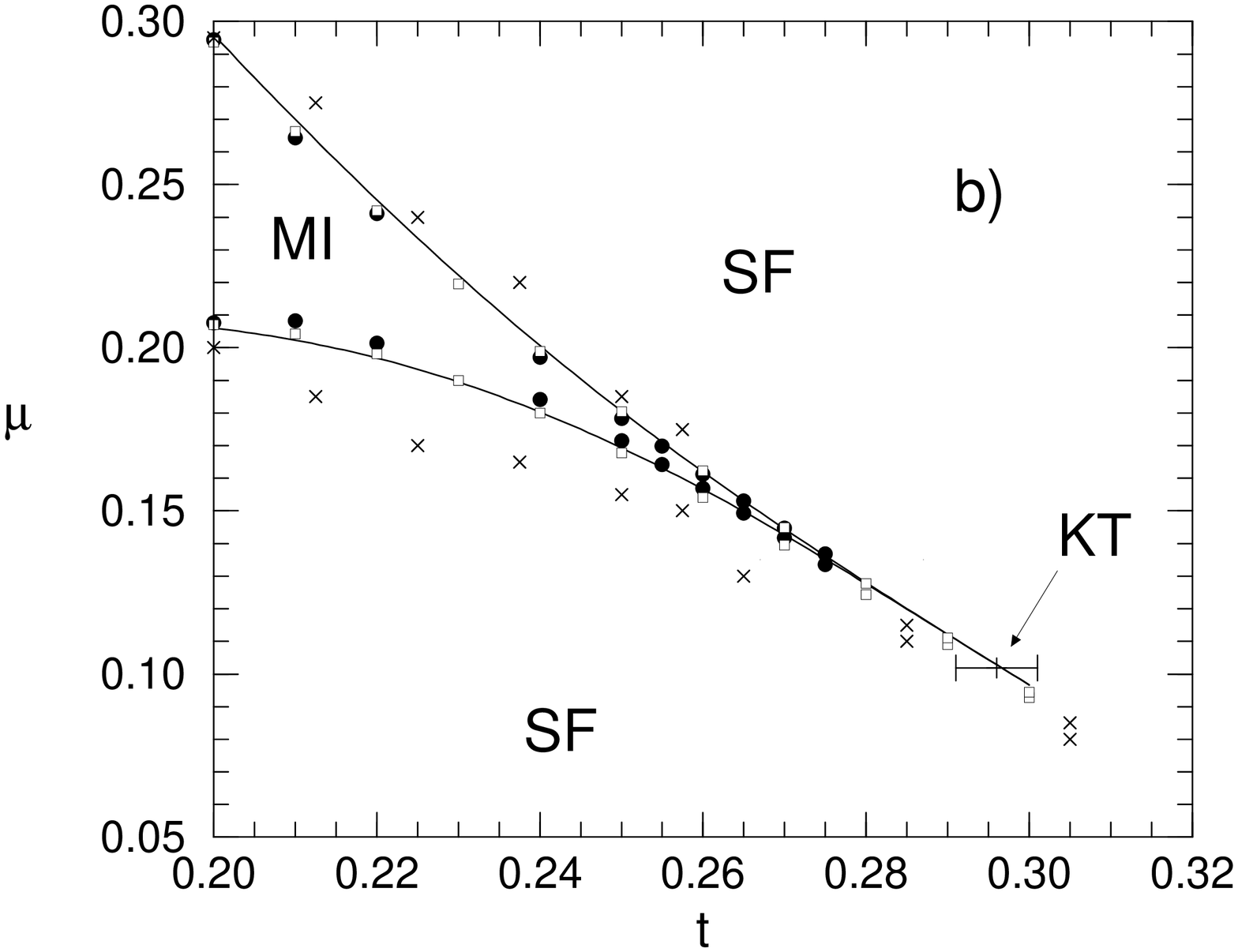,
      height=\figheight,width=\figwidth,angle=-90}
    \protect\caption[]{  Same as \ref{Fig.phasediagram.MI.noU1},
	with the tip of the Mott-insulator on an expanded scale.
	}
    \label{Fig.phasediagram.MI.noU1.expanded}
  \end{center}
\end{figure}

Using the methods described above we determine the phase diagram in the
presence of on-site interactions only. Fig. \ref{Fig.phasediagram.MI.noU1}
shows the 
Mott-insulator with density $\rho=1$, surrounded by the superfluid phase.
In Fig. \ref{Fig.phasediagram.MI.noU1.expanded} the tip of the insulator
is shown on an expanded scale.
The very pointed tip of the insulating region reflects the closing of the 
energy gap given by Eq. (\ref{KT_behaviour}). 
Fig. \ref{Fig.phasediagram.MI.noU1} also shows results for the 
commensurate-incommensurate phase transition from twelfth order 
perturbation theory. The excellent agreement with DMRG confirms the high 
accuracy achieved.

\begin{figure}[t]
  \begin{center}
    \protect\epsfig{file=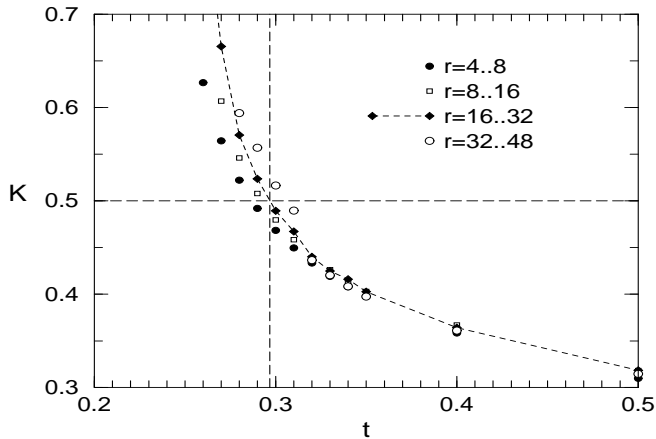,
      height=\figheight,width=\figwidth,angle=-90}
    \protect\caption[]
    {	$K$ vs $t$ at the Kosterlitz-Thouless transition from the
 	density one Mott-insulator to the superfluid phase. The critical point
	is determined by finding the $t$ at which $K=1/2$. The biggest system
	sizes are $L=256$ sites except for $t=0.29 , 0.3 , 0.31$, where 
	they are $L=1024$.
    }
    \label{Fig.K_vs_t.noU1.rho1}
  \end{center}
\end{figure}

\begin{table}[b]
   \caption{ 
	The location of the critical point $t_c$ depending on the
	interval of $r$ used for the fits to 
	$\Gamma(r) = \langle b^\dagger_r b_0 \rangle$.
   }	
   \begin{tabular}{cc}
	$r$   	&   	$t_c$ \\
	\hline
	$4 \leq r \leq 8 $	&	$0.2874 \pm 0.0001$\\	
	$8 \leq r \leq 16$	&	$0.2938 \pm 0.0001$\\
	$16 \leq r \leq 32$	&	$0.2968 \pm 0.0003$\\
	$32 \leq r \leq 48$	&	$0.3062 \pm 0.0003$\\
	$48 \leq r \leq 64$	&	$0.3107 \pm 0.01\phantom{00}$ 
   \end{tabular}
   \label{Table.1}
\end{table}

To find the critical point of the Kosterlitz-Thouless transition at the tip
of the Mott-insulator, we determine the $t$ at which $K=1/2$. 
As described in Section \ref{K_extrapolation},
the $K$ are determined by fitting power-law behavior to the decay of the
hopping correlation function $\Gamma(r) = \langle b^\dagger_r b_0 \rangle$.
Due to finite-size effects, the result can depend on the interval of $r$ that
is used for this fit. While we found $16 \leq r \leq 32$ to be reasonable 
in general,
there are logarithmic finite-size effects at the Kosterlitz-Thouless transition
that require a more detailed inspection. Fig. \ref{Fig.K_vs_t.noU1.rho1} 
shows $K$'s that were determined with different fitting intervals 
plotted versus $t$.

For the $t=0.29$, $t=0.3$ and $t=0.31$ systems with $L=512$ and $L=1024$
sites were calculated to keep finite-size effects small. 
Table \ref{Table.1} shows the $t_c$ found with different fitting intervals.
Due to the finite-size effects the $t_c$ go to higher $t$ as the 
fitting interval is shifted to bigger $r$. While using 
a fitting interval of $16 \leq r \leq 32$ is
a good compromise between avoiding finite-size effects by using small $r$, 
and finding an asymptotic value for the decay by using big $r$, the
error determined by this fit alone underestimates the true error.
With an error estimated from the effect of the choice of the fitting interval,
we find $t_c=0.297 \pm 0.01$.

Determination of the Kosterlitz-Thouless transition was attempted 
in several previous studies. 
For a truncated model with a maximum of two particles per site
the critical point was found at \mbox{$t_c^{BA}=1/(2 \sqrt{3}) \approx 0.289$}
with the Bethe-Ansatz\cite{Krauth:1991}. 
In a combination of exact-diagonalization
for system with up to $L=12$ sites and renormalization group Kashurnikov and Svistunov found \mbox{$t_c = 0.304 \pm 0.002$\cite{Kashurnikov:1996}}, 
and together with Kravasin found 
\mbox{$t_c=0.300\pm0.005$\cite{Kashurnikov:1996.2}} in a quantum Monte Carlo
study.  
An exact diagonalization approach reported the critical point to
be at \mbox{$t_c = 0.275 \pm 0.005$\cite{Elesin:1994},} 
and a study using 12th order strong coupling 
expansions\cite{ElstnerMonien:1999} located it at \mbox{$t_c = 0.26 \pm 0.01$.}

There were two previous studies using DMRG to locate the Kosterlitz-Thouless
transition in the Bose-Hubbard model. In the first study\cite{Pai:1996}
a particle cut-off of $n=4$ per site  was used, and the critical point was 
determined by using a phase twist $\phi = \pi$ and periodic boundaries 
to calculate the superfluid stiffness. The critical point was found at 
\mbox{$t_c = 0.298$}. In another DMRG study\cite{KuehnerMonien:1998},
the critical point was found at \mbox{$t_c = 0.277 \pm 0.01$} by 
determining the $t$ at which $K=1/2$, similar to the procedure used in
the present work. In both of these DMRG studies the numerical accuracy 
was limited due to the use of the infinite size version
of the DMRG algorithm, periodic boundary conditions and a small number
of states in the DMRG basis. The biggest system sizes were smaller than $L=80$.

The range of these results demonstrates the difficulty involved in 
determining the critical point of the Kosterlitz-Thouless transition,
which is mostly due to logarithmic finite-size effects close to 
the critical point. The large system
sizes used in this paper should compensate for this within the given
error bars, and yield a reliable result.

\begin{figure}[t]
  \begin{center}
    \epsfig{file={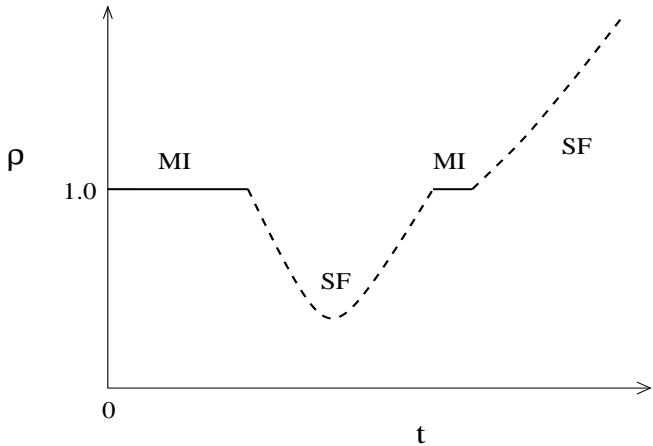}, height=\fignorheight,width=\fignorwidth, angle=0}
    \caption 
    { 
	Illustration of the phase transitions between the Mott-insulator (MI)
    	and the superfluid phase (SF) on a line of 
	constant chemical potential $\mu = 0.15$. 
    }
	\label{reentrance_ill}
  \end{center}
\end{figure}

The phase diagram shown in Fig. \ref{Fig.phasediagram.MI.noU1} has 
a very interesting feature. Imagine moving on a line of 
constant chemical potential $\mu$, for example $\mu=0.15$, and
starting at small $t$, moving toward bigger $t$. The particle density 
along this line is illustrated in Fig. \ref{reentrance_ill}. 
For small $t$ the system is in the Mott-insulator phase. 
At $t \approx 0.1$ there is a phase transition
to the superfluid phase, with densities $\rho<1$. The density decreases
up to a minimum, then it start increasing again. At $t \approx 0.26$
the density goes up to $\rho=1$ again, and there is another phase transition,
this time reentering the Mott-insulating phase from the superfluid phase.
Increasing $t$ further leads to another phase transition from
the Mott-insulator to the superfluid phase, this time with $\rho>1$, and
the density increasing further with increasing $t$. 

\begin{figure}[b]
  \begin{center}
    \epsfig{file={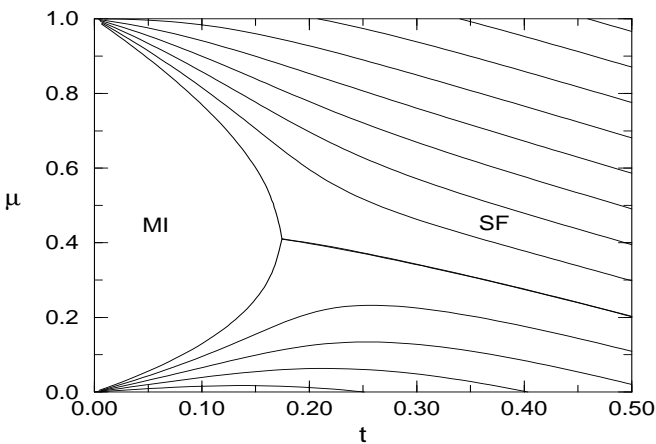}, height=\figheight,width=\figwidth, angle=-90}
    \caption 
    { 
	The Mean-Field phase diagram\cite{Ramakrishnan:1993} in 
	dimensionless units, showing lines of constant density.
    }
	\label{MeanFieldPhasediagram}
  \end{center}
\end{figure}

To gain more insight into this, we compare to results from a mean field 
approach\cite{Ramakrishnan:1993}. Fig. \ref{MeanFieldPhasediagram}
shows the phase diagram with the Mott-insulator at density $\rho=1$,
surrounded by the superfluid phase. In the superfluid phase,
the lines of constant density slope downward as $t$ is increased. 
This is not only found in one dimension, but in all dimensions.
The limit of $t \rightarrow \infty$
corresponds to keeping $t$ constant and setting the interactions to zero. 
If the interactions are zero the system goes from a superfluid phase to a
Bose-Einstein condensate, in which every particle has an energy of $-2 t$.
If the chemical potential is smaller than $-2 t$, the system is empty, 
because it costs energy to put a particle in, and for chemical potentials 
bigger than $-2 t$ the number of particles goes to infinity, because every 
additional particle reduces the total energy of the system. 
Going back to the picture of constant interactions and changing $t$, 
this means that the density of the system always goes to infinity as $t$ 
is increased. 

\begin{figure}[tb]
  \begin{center}
    \epsfig{file={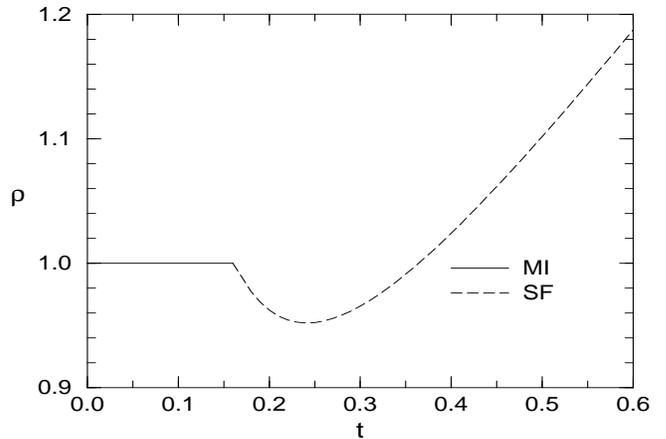}, height=\figheight,width=\figwidth, angle=-90}
    \caption 
    { The density on a line of constant chemical potential $\mu = 0.3$ in the Mean-Field phase diagram (Fig. \ref{MeanFieldPhasediagram}). }
	\label{MeanFieldPhasediagram.cut}
  \end{center}
\end{figure}

In dimensions two and higher the superfluid-insulator transition on the line
of constant density is a second order transition, and the tip of the 
insulating region is round. Fig. \ref{MeanFieldPhasediagram.cut} shows
the density on a line of constant chemical potential $\mu=0.3$. At the phase
transition from the Mott-insulator to the superfluid phase the density first
drops as $t$ is increased, and then increases again. 

In one dimension the tip of the insulating region is very long and narrow due 
to the Kosterlitz-Thouless transition, and it is possible to reenter into the 
insulator at $\rho=1$.

\section{Nearest-neighbor interaction}
\label{NearestNeighborInteraction}

Longer ranger interactions have been found to be important in 
experiments\cite{Haviland:1998,Oudenaarden:1996,Oudenaarden:1998}. We now 
include nearest-neighbor interactions by setting $V=0.4$. 
Due to the nearest-neighbor interactions a new insulator phase appears 
at half integer densities. 
It is a charge density wave phase (CDW) with a wavelength of two sites,
and like the Mott-insulator at integer density it has an excitation gap 
and is incompressible. The crystalline order is characterized by a 
non-zero structure factor
\begin{equation}
 S_\pi=\frac{1}{N^2} \sum_{ij} (-1)^{\mid i-j \mid} 
 \langle n_i n_{j}\rangle \;.
\label{StructureFactor}
\end{equation}
In Fig. \ref{Densityfluctuations.rho0.5} the local density oscillations in the
charge density wave phase are shown. A small boundary effect can be seen, 
but the main feature are long-range density oscillations  throughout the 
system that do not decay. An order parameter $\langle n_i - \rho \rangle$ 
can be defined to describe this charge density wave, even in one dimension.

In one dimension the superfluid phase is signalled by a diverging
correlation length 
\begin{equation}
  \label{correlation_length}
	\xi^2 = \sum_r r^2   \langle b^\dagger_r b_0 \rangle 
	/ \sum_r  \langle b^\dagger_r b_0 \rangle \;,
\end{equation}
and a non-zero superfluid stiffness
\begin{equation}
	\rho_s = \lim_{\phi \rightarrow 0} L \frac{\partial^2 E_0(\phi,L)}
						{\partial \phi^2}\;,
\label{rhos}
\end{equation}
which is proportional to the Drude weight $D \sim \rho_s$. In two and higher
dimensions there is also an order parameter 
$\langle b^\dagger_i \rangle \neq 0$.
In one dimension the whole superfluid phase is critical,  and there
is no order parameter.

At the transition from the charge density wave phase to the
superfluid phase both types of order 
are involved: the crystalline order in the charge density wave phase and the 
superfluid order in the superfluid phase. In addition to a direct 
phase transition from the charge density wave to the superfluid at
which the crystalline order vanishes at the same point where the
superfluid order appears, there is the possibility of an intermediate 
phase. Tab. \ref{Possible_phases} shows the possible phases close to
density $\rho=1/2$ in a bosonic system with on-site and nearest-neighbor
interaction in two or higher dimensions. In addition to the charge density
wave and the superfluid phase, supersolids that have both forms of 
order were found in two dimensional models\cite{Otterlo:1994,Batrouni:1995}.
Baltin and Wagenblast\cite{BaltinWagenblast:1997} found a region that has
neither superfluid stiffness nor charge density wave in a one dimensional 
bosonic model in the high density. The possible existence of such a phase 
was also recently predicted for a two dimensional bosonic model in the 
high density limit by Das and Doniach\cite{DasDoniach:1999}, who call it 
a Bose-metal.

\begin{table}[b]
\caption{Possible phases and their order parameters close to density
	$\rho=1/2$.}
\begin{tabular}{ccc}
$S_\pi$	&	$\rho_s$ 	&	 phase\\
\hline
$\neq 0$&	$=0$		&	 charge density wave\\
$=0$	&$\neq 0$		&	 superfluid\\
$\neq 0$&$\neq 0$		&	 supersolid\\
$=0$	&	$=0$		& 	 Bose-metal\\
\end{tabular}
\label{Possible_phases}
\end{table}

In Appendix \ref{StrongCouplingExpansion} strong coupling expansions
are used to illustrate the difference between the 
commensurate-incommensurate phase transition at $\rho=1/2$  
in one and two dimensions. Strong coupling expansions can be used
to study the insulator, but not the superfluid. To study the low-energy behavior of the superfluid phase the Luttinger liquid can be used. 
In addition to the basic Luttinger liquid Hamiltonian
(Eq. (\ref{LuttingerHamiltonian})), the lattice and the interactions 
introduce scattering terms (Eq. (\ref{UmklappTerm})). These only contribute 
at $\rho=1/2$, where they can drive the system into a different phase, 
but not at nearby densities. At incommensurate densities close to
$\rho = 1/2$, the wavefunction is incommensurate with the lattice and 
hence cannot be pinned to the lattice to form an insulator. Of course
this would be changed if there were impurities or disorder, but a
pure system in one dimension is in the Luttinger liquid phase unless 
it is at a density commensurate with the lattice and the interactions. 

At density $\rho=1/2$ DMRG can be used to determine if there is an 
intermediate phase or a direct phase transition from the charge density 
wave phase to the superfluid phase. To do this, we investigate the 
relationship between the superfluid and crystalline order at the phase 
transition. The onset of superfluidity is signaled by a diverging correlation 
length $\xi$ (Eq. (\ref{correlation_length})), charge density order is
meassured by $S_\pi > 0$ (Eq. (\ref{StructureFactor})).

\begin{figure}[bt]
  \begin{center}	
    \epsfig{file={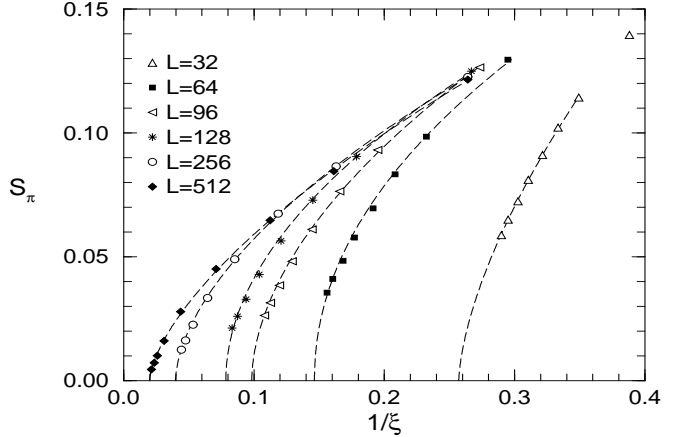},height=\figheight,width=\figwidth,angle=-90}
 \caption 
    {	Structure factor $S_\pi$ versus the inverse correlation length $\xi$ for 
      	different system sizes. The dashed lines show power-law fits.
}
\label{S_versus_Xi}
\end{center}			
\end{figure}

The model is in the universality class of the xy-model at the phase
transition on the line of constant density at $\rho=1/2$. Due to
the Kosterlitz-Thouless transition expected on this line, 
it is difficult to determine exactly when the structure factor or 
the inverse correlation length go to zero.
But it is possible to study the dependence of the structure factor on
the correlation length. Fig. \ref{S_versus_Xi} shows 
that for small values the structure factor depends on the correlation length
by a power-law:
\begin{equation}
	\xi(S_\pi)^{-1} - \xi(S_\pi=0)^{-1} \sim  (S_\pi)^\alpha \; .
\end{equation}
To keep the effect of the boundaries small for the calculation of both 
the structure factor $S_\pi$ and the correlation length $\xi$, only sites that
were at least a quarter of the system size away from the boundaries
were taken into account.

\begin{figure}[tb]
  \begin{center}	
    \epsfig{file={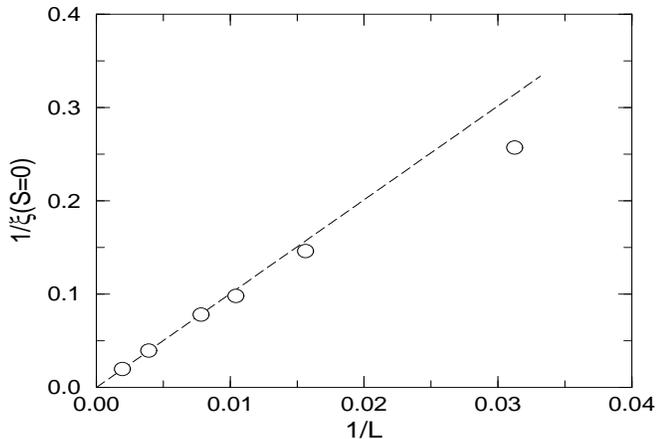},height=\figheight,width=\figwidth,angle=-90}
 \caption 
    { 	The inverse correlation length $1/\xi$ at $S_\pi=0$ obtained from 
	Fig. \ref{S_versus_Xi} versus the inverse system size. 
	The dashed line shows a linear fit to the three biggest system sizes. 
}
\label{Xi0_vs_1dL}
\end{center}			
\end{figure}

In Fig. \ref{Xi0_vs_1dL} the extrapolated inverse correlation length 
$1/\xi(S_\pi=0)$ at zero structure factor is plotted against the inverse 
system size. The linear fit to the three biggest system sizes shows that 
$1/\xi(S_\pi=0)$ goes to zero for infinite systems. From this and the
power-law behavior in Fig. \ref{S_versus_Xi} we conclude that there is 
a power-law dependence of the structure
factor on the correlation length, and that in infinite systems the correlation 
length diverges at the same point at which the structure factor goes to
zero. This means that there is a direct phase transition from the charge
density wave phase to the superfluid phase, and no supersolid or normal
phase in between.

\begin{figure}[b]
  \begin{center}	
    \epsfig{file=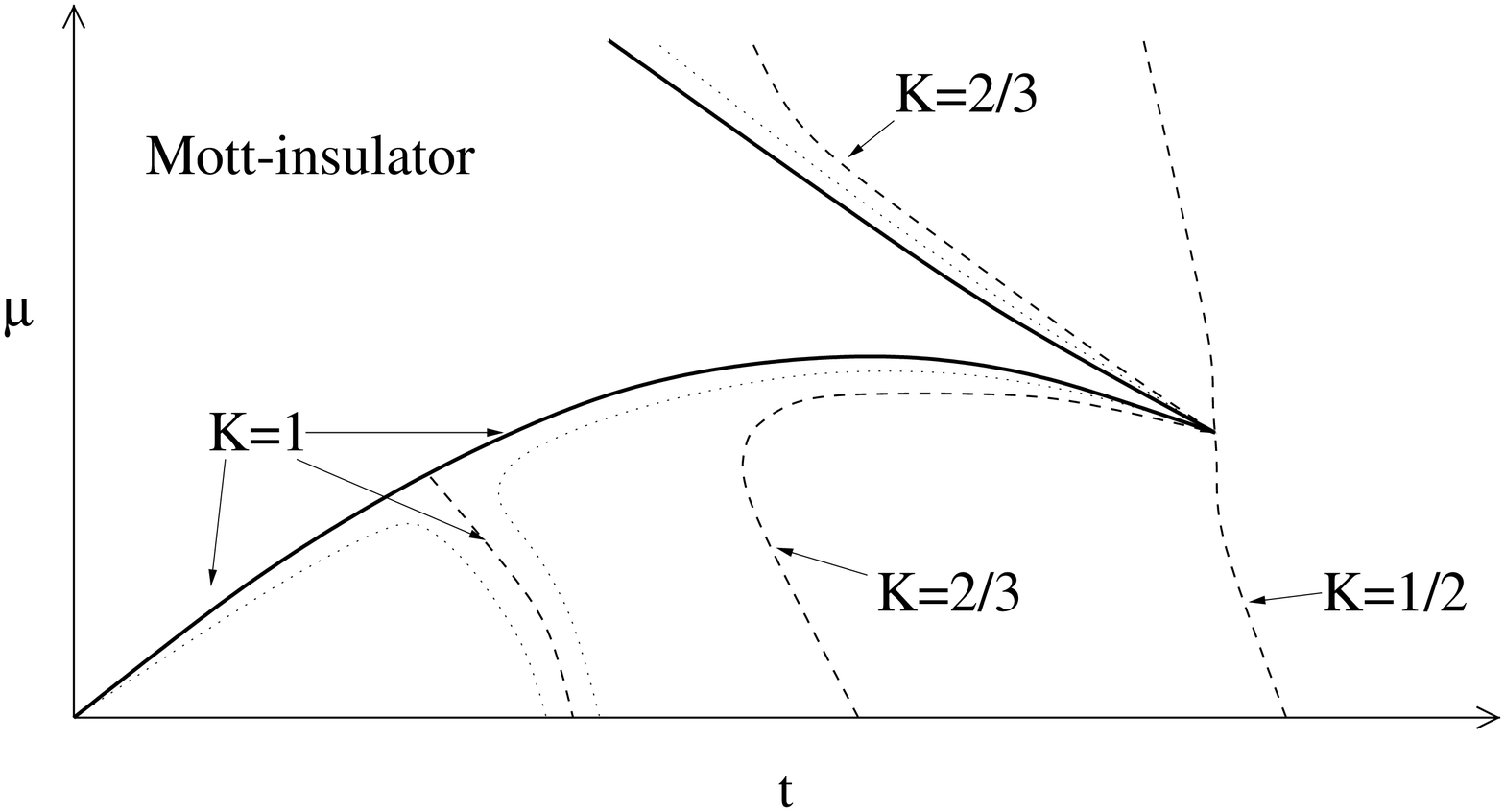,height=4.5cm,width=\fignorwidth,angle=0}
     \caption
     {
	Illustration of the lines of constant $K$ in the phase diagram.
	The dotted lines indicate $K$ that are slightly smaller or bigger
	than $K=1$.
     }	
	\label{K_illustration}		
 \end{center}
\end{figure}
The phase boundaries of the charge density wave phase can be found in
the same way as those of the Mott-insulator, and we use the methods
used for the on-site only interaction case to calculate the
phase diagram. Fig. \ref{Fig.Phasediagram.U1}
shows the phase diagram in the region of the $\rho=1/2$ charge 
density wave phase and the $\rho=1$ Mott-insulator.
Like the shape of the Mott-insulator with on-site interaction only, the 
shapes of the insulating 
regions reflect the Kosterlitz-Thouless transitions at the tips. The tips
are also bending down, allowing re-entrance phase transitions from the 
superfluid to the insulating phases. We find the Kosterlitz-Thouless 
transitions at $t_c^{MI} = 0.404 \pm 0.02$ for the Mott-insulator, and at
$t_c^{CDW} = 0.125 \pm 0.003$ for the charge density wave phase.
The critical point at the tip of the charge density wave phase had been 
found in a quantum Monte Carlo study at \mbox{$t \approx 0.1$}.
The accuracy of  $t_c^{MI}$ is relatively low because $K$ only changes 
very slowly if $t$ is changed close to this transition. This had already
been observed in an earlier DMRG study\cite{KuehnerMonien:1998}, 
where the critical points were 
found at  \mbox{$t_c^{CDW} =0.118 \pm 0.004$} and 
\mbox{$t_c^{MI} \approx0.325 \pm 0.05$}. That study used the infinite 
size version of the DMRG, and the biggest systems were $L=76$ 
sites long. Since there are logarithmic finite-size effects at the
Kosterlitz-Thouless transitions, the values determined in the present work,
where systems with up to $L=512$ sites are used, are much more accurate.

At the phase boundaries of the charge density wave phase the 
Luttinger Liquid parameter is $K=4$ except for the  Kosterlitz-Thouless
transition at the tip, where it is $K=2$. The charge density wave phase is 
surrounded by a region where $K>1$. 
The effective interactions in the Luttinger-Liquid are attractive 
for $K<1$ and repulsive for $K>1$. Kane and Fisher showed that in 
the repulsive region a single impurity turns the system into an 
insulator\cite{KaneFisher:1992,GlazmanLarkin:1997}. An even larger 
region of the Luttinger liquid is driven into a glass phase if $K>2/3$
for any finite quenched disorder\cite{GiamarchiSchulz:1987,Fisher:1989}.

In the phase diagram these regions ($K>1$ and $K>2/3$) are determined 
by doing calculations for different densities and $t$. For example,
to determine the $t$ for which $K(t)=1$ at a given density $\rho$, we
do calculations for different $t$ until we find a pair of $t_1$ and $t_2$
that are close to each other, and $K(t_1)>1>K(t_2)$. We then determine
$t(K=1)$ by linear interpolation. To find the boundaries of the repulsive 
Luttinger Liquid region, we calculate $t(K=1)$ for various densities.

The lines with $K=1$ and $K=2/3$ are shown in Fig. 
\ref{Fig.Phasediagram.U1}. The repulsive Luttinger Liquid ($K>1$) region 
completely surrounds the charge density wave phase. Instead of going 
to $t=0$ as the Mott-insulator
is approached, the $K=1$ line ends in the side of the Mott-insulating region,
where $K=1$ for all of the commensurate-incommensurate phase boundary.
Fig. \ref{K_illustration} illustrates how the two lines with $K=1$ meet
at the phase boundary of the Mott-insulator. Although we could not obtain more
detailed results for densities closer to $\rho=1$, we argue that
the lines with $K>1$ bend towards $t=0$ as the density gets closer to one, 
while those with $1/2<K<1$ bend towards the tip of the Mott-insulator.

Lines of constant $K$ with $1/2<K<1$ are discontinuous at $\rho=1$, where
the system is an insulator for $K>1/2$, with the tipped shape reflecting the
Kosterlitz-Thouless behavior. Lines with $K \leq 1/2$ are round at $\rho=1$,
and do not reflect the Kosterlitz-Thouless behavior.
Analogous to $\rho=1$, we also observe that the lines of constant $K$ 
are round for $K\leq 2$ and $\rho=1/2$, where the charge density wave 
phase ends in a Kosterlitz-Thouless transition at $K=2$.

\onecolumn{
\begin{figure}[bt]
 \begin{center}
\epsfig{file=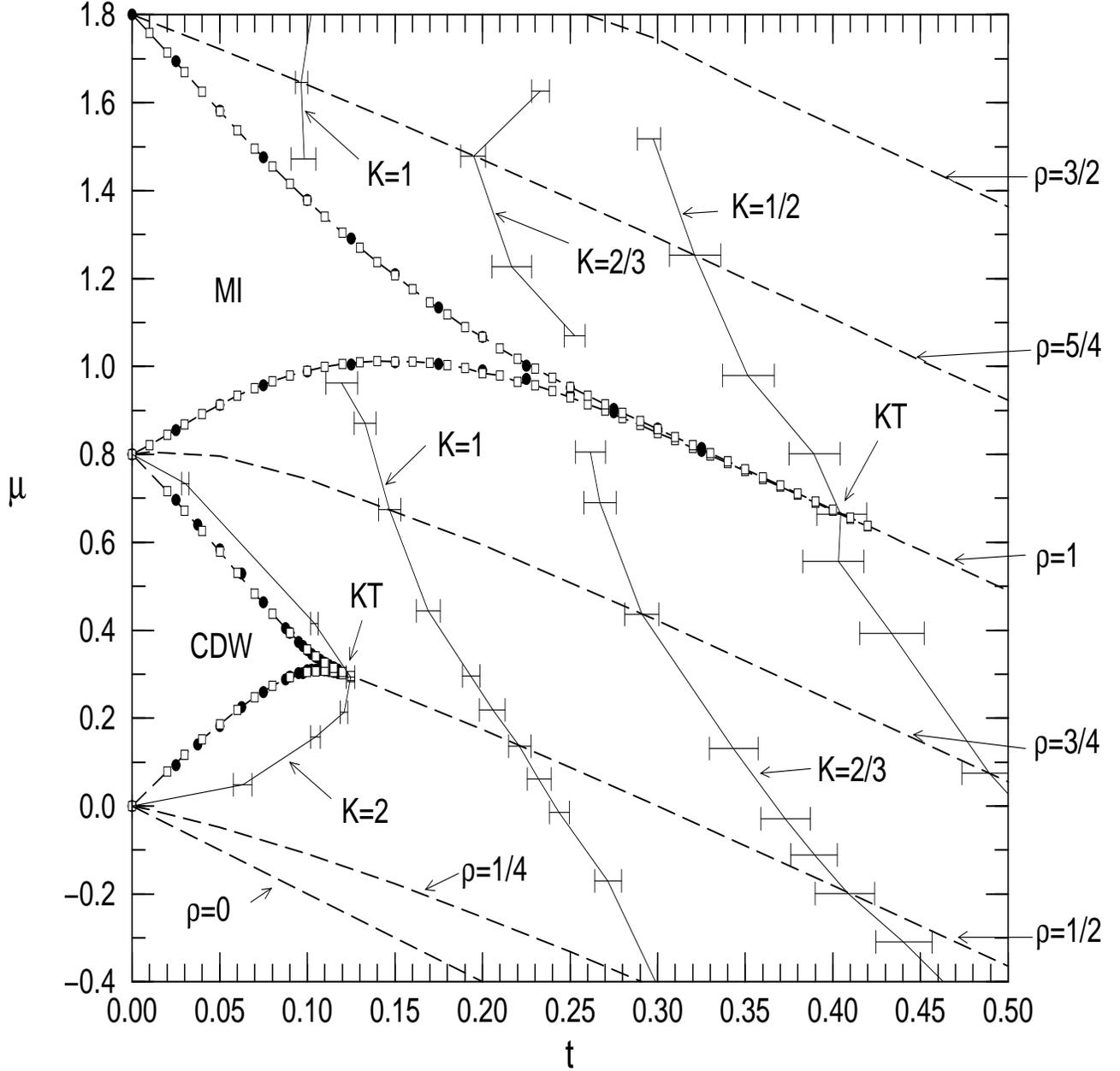,height=17.2cm,width=17.2cm,angle=-90}
   \caption
	{
	The phase diagram of the Bose-Hubbard model with nearest-neighbor 
	interaction $V=0.4$. The Mott-insulator (MI) at density $\rho=1$ 
	and the charge density wave phase (CDW) at density $\rho=1/2$
	are surrounded by the superfluid phase.
	The long-dashed lines show the lines of constant density. The
	solid lines are lines of constant $K$.
	The $K=2$ line crosses the density $\rho=1/2$ line at the 
        Kosterlitz-Thouless (KT) transition at the tip of the charge density 
	wave phase.
	In the region left of the $K=1$ line, where $K>1$,
	the superfluid phase is turned into an insulator by a single
	impurity\cite{KaneFisher:1992}. In the presence of disorder
	the region left of $K=2/3$ line is turned into a Bose-glass 
	phase\cite{GiamarchiSchulz:1987,Fisher:1989}. 
	The Kosterlitz-Thouless (KT) transition
	at the tip of the Mott-insulator is at the point where the $K=1/2$ 
	line line intersects the $\rho=1$ line.
	}
    \label{Fig.Phasediagram.U1}
  \end{center}
\end{figure}
\twocolumn

\section{ac-conductivity}
\label{ac-conductivity}

The repulsive region of the Luttinger liquid is turned into an insulator
by a single impurity\cite{KaneFisher:1992}. This raises 
the question if there is a qualitative difference between the 
conductivity in the repulsive region and the attractive 
region of the Luttinger liquid in the pure system. 
The regular part of the conductivity is given by:
\begin{eqnarray}
\nonumber
        \sigma_1^{reg}(\omega) 
	&=& \frac{1}{L} \sum_{m\neq0} 
		\frac{{| \langle m \mid j_{q=0} \mid 0 \rangle |}^2}
                     {E_m - E_0} \delta{(\omega- (E_m - E_0))}\\
\nonumber
	&=& - \frac{1}{\omega \, \pi \, L} 
		\text{Im} \lim_{\eta \rightarrow 0^{+}} \\
	&&	\langle 0 \mid j^{\dagger}_{q=0}    
			\frac{1}{\omega + E_0 - H + i \eta}
			j_{q=0} \mid 0 \rangle \;,
\end{eqnarray}
and the current operator is 
\begin{equation}
\label{currentop}
        j_q = i t \sum_{n} e^{-i q n} (b^\dagger_{n+1} b_{n} 
                                        - h.c. ) \;.
\end{equation}

Recent developments with DMRG make the calculation of dynamical correlation
functions like the ac-conductivity possible\cite{KuehnerWhite:Dynamics}. 
The conductivity at a frequency $z=\omega + i \eta$ can be calculated 
as the direct product of the current operator applied to the groundstate
\begin{equation}
	\mid j_{q=0} \rangle = j_{q=0} \mid 0 \rangle \;,
\end{equation}
and a correction vector
\begin{equation}
	\mid x(z) \rangle = \frac{1}{\omega + E_0 - H + i \eta} 
				\mid j_{q=0} \rangle \;.
\end{equation}
By using these two states and the groundstate $| 0 \rangle$ 
as DMRG target states, the conductivity $\sigma_1^{reg}(z)$ can be
calculated very precisely. To calculate the conductivity over an interval
of width $\eta$ ranging from $\omega_1$ to $\omega_2$, we use correction 
vectors $\mid x(\omega_1+i \eta) \rangle$ and $\mid x(\omega_2+i \eta) \rangle$
as target states. At the end of the DMRG calculation, when
the DMRG basis is optimized to represent these states, we calculate 
the conductivity from $\omega_1$ to $\omega_2$. Repeating this 
procedure for neighboring intervals, we piece together the conductivity for
a whole range of frequencies.

The finite broadening $\eta$ in the correction vectors is only used to
obtain appropriate DMRG target states. To calculate the spectrum within
the DMRG basis, we use a Lanczos method that yields approximate
eigenstates of the Hamiltonian. The broadening used in our plots is then
applied to the discrete peaks found with the Lanczos method, and is only
used for better visualization.

DMRG calculations work best with open boundary conditions. How the current
operator is applied in a system with open boundary conditions is discussed
in \mbox{Appendix \ref{Appendix:currentop}}.

\begin{figure}[b]
  \begin{center}
   \epsfig{file={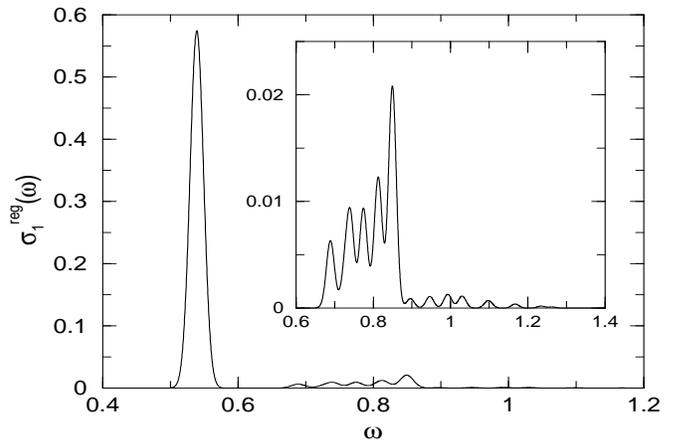}, height=\figheight,width=\figwidth, angle=-90}
    \caption
    {
	The conductivity $\sigma_1^{reg}(\omega)$ in the Mott-insulator
	at density $\rho=1$, $t=0.05$, $U=1$, $V=0.4$, system size
	$L=256$, $m=128$ states,
	two correction vectors as target states with $\eta=0.05$,
	and broadening $\eta_g=0.01$ for the plot. The inset shows the 
	same data on an expanded scale.
   }
   \label{conductivity.MI}
  \end{center}
\end{figure}

The conductivity in the Mott-insulator phase is shown in 
Fig. \ref{conductivity.MI}. There is an energy gap of 
\mbox{$\Delta \omega = 0.54$}, with a big peak after that and only small 
excitations at higher energies. Since it is an insulator phase, we expect 
to find no Drude weight. With the kinetic energy defined as
\begin{equation}
	\langle T \rangle = t \sum_{n} \langle b^\dagger_{n+1} b_{n} 
                                       + h.c. \rangle \;,
\end{equation}

the Drude weight is given by
\begin{equation}
	D = - \frac{1}{L} \langle T \rangle 
            -  2\int d\omega \; \sigma_1^{reg}(\omega)\;.
\end{equation}
Note that the Drude weight is proportional to the superfluid 
stiffness $\rho_s$ given in Eq. (\ref{rhos}). 

Since the kinetic energy $\langle T \rangle$ can be calculated directly 
with DMRG, and we expect $D=0$, this is an opportunity to verify the
consistency of the calculation. In Tab. \ref{Drudew.MI}  the Drude weight
is shown for various system sizes. A small finite-size effect can be
seen in the data. From the differences in the individual values 
we estimate the error of the Drude weight $\Delta D = 0.02 t$, 
or $2\%$ of $- \langle T \rangle/L$.

\begin{table}[t]
\caption{The Drude weight $D$ for different system sizes in 
	the Mott-insulator with density $\rho=1$, $t=0.05$,
	$U=1$,$V=0.4$. $\langle T \rangle$ is the kinetic energy,
	$\int \sigma_1^{reg}(\omega) d\omega$ the integral over
	the ac-conductivity. Also shown is the number of states $m$, 
	the broadening $\eta$ of the correction vectors, 
	and the truncation error $\Delta$.
	}
\begin{tabular}{crcclcc}
	$L$ 	&$D/t$	&$\langle T \rangle / (t L)$	
				&$2/t \int \sigma_1^{reg}(\omega) d\omega$	&$\eta$	&$m$	&$\Delta$\\
\hline
	32	&0.0084	&0.6392		&0.6307		&0.05	&128	&$10^{-7}$\\
	64	&-0.0040&0.6392		&0.6432		&0.05	&128	&$10^{-7}$\\
	128	&-0.0075&0.6392		&0.6467		&0.05	&128	&$10^{-7}$\\
	256	&-0.0036&0.6392		&0.6428		&0.05	&128	&$10^{-7}$\\
\end{tabular}
\label{Drudew.MI}
\end{table}

\begin{figure}[b]
  \begin{center}
   \epsfig{file={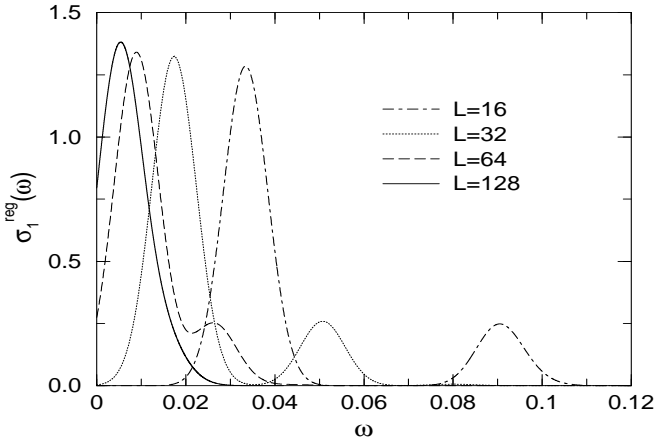}, height=\figheight,width=\figwidth, angle=-90}
    \caption
    {
	The precursor peaks in the conductivity $\sigma_1^{reg}(\omega)$ 
	in the repulsive region of the superfluid phase at density $\rho=3/4$, 
	$t=0.05$, $U=1$, $V=0.4$, system size $L=128$ and $m=128$ states.
	Broadening $\eta=0.02$ for correction vectors,
	and $\eta_g=0.005$ for the plot. 
   }
   \label{precursor.SF.RLL}
  \end{center}
\end{figure}
In the superfluid phase we find precursor peaks at small frequencies in 
the conductivity. They are due to the finite width\cite{KuehnerWhite:Dynamics}
of the wavevector $q$, which is $\Delta q = 4 \sqrt{3}/L$. 
Fig. \ref{precursor.SF.RLL} shows these precursor peaks in the repulsive 
Luttinger liquid for different system sizes. As the system size
is increased the precursor peaks move towards $\omega=0$. For the
calculation of the Drude weight these peaks should not contribute, and
we use low-energy cut-offs to ignore them.

\begin{figure}[t]
  \begin{center}
   \epsfig{file={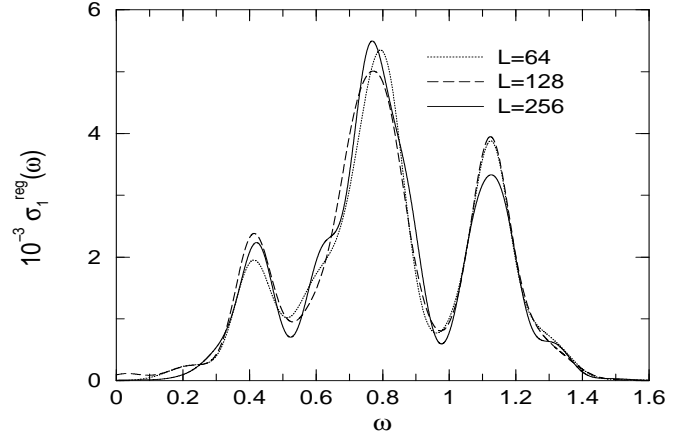}, height=\figheight,width=\figwidth, angle=-90}
    \caption
    {
	The conductivity $\sigma_1^{reg}(\omega)$ in the repulsive 
	region of the superfluid phase at density $\rho=3/4$, 
	$t=0.05$, $U=1$, $V=0.4$. Data shown is for different 
	system sizes with ($L=64$,$m=256$,$\eta=0.05$), 
	($L=128$,$m=256$,$\eta=0.05$) and ($L=256$,$m=512$,$\eta=0.2$).
	Broadening $\eta_g=0.05$ for the plot, and the low-frequency 
	cut-offs given in Tab.\ref{Drudew.SF.RLL}.
   }
   \label{conductivity.SF.RLL}
  \end{center}
\end{figure}

\begin{figure}[b]
  \begin{center}
   \epsfig{file={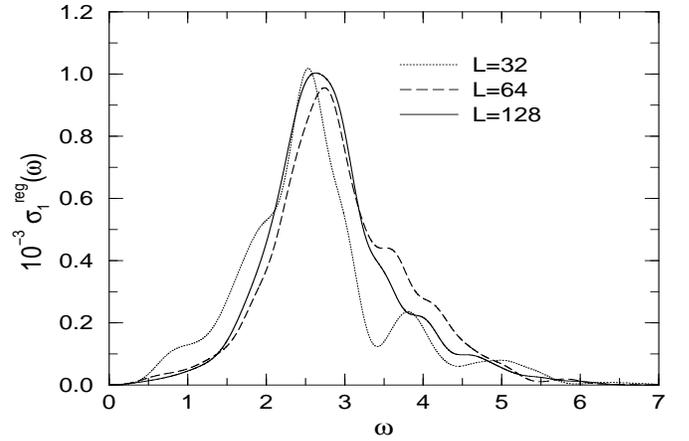}, height=\figheight,width=\figwidth, angle=-90}
    \caption
    {
	The conductivity $\sigma_1^{reg}(\omega)$ in the attractive
	region of the superfluid phase at density $\rho=3/4$, $t=0.5$, 
	$U=1$, $V=0.4$. System size $L=32$ and $L=64$ with $m=128$ states,
	and $L=128$ with $m=256$ states.
	Broadening $\eta=0.2$ for the correction vectors,
	$\eta_g=0.2$ for the plot, and the low-frequency cut-offs given 
	in Tab.	\ref{Drudew.SF.ALL}.
   }
\label{conductivity.SF.attractive}
  \end{center}
\end{figure}

\begin{table}[t]
\caption{The Drude weight $D$ for different system sizes in the repulsive 
	region of the superfluid phase at density $\rho=3/4$, $t=0.05$,
	$U=1$,$V=0.4$. Same notation as in Tab.\ref{Drudew.MI}.
	}
\begin{tabular}{cccccccl}
	$L$ 	&$D/t$	&$\langle T \rangle / (t L)$
				&$2/t \int \sigma_1^{reg}(\omega) d\omega$
							&$\eta$	&$m$	&$\Delta$	&$\omega_{c}$\\
\hline
16	&0.776	&0.874	&0.099	&0.2	&128	&$5\times10^{-7}$&0.2\\
32	&0.791 &0.885	&0.094	&0.2	&128	&$7\times10^{-7}$&0.1\\
64	&0.799	&0.890	&0.091	&0.2	&128	&$5\times10^{-6}$&0.08\\
64	&0.798	&0.890	&0.092	&0.05	&256	&$1\times10^{-6}$&0.08\\
128	&0.799	&0.892	&0.093	&0.2	&128	&$7\times10^{-6}$&0.02\\
128	&0.801	&0.892	&0.091	&0.1	&256	&$6\times10^{-7}$&0.02\\
128	&0.793	&0.892	&0.099	&0.05	&128	&$1\times10^{-5}$&0.02\\
128	&0.797	&0.892	&0.095	&0.05	&256	&$1\times10^{-6}$&0.02\\
256	&0.800	&0.893	&0.093	&0.2	&512	&$5\times10^{-7}$&0.02\\
\end{tabular}
\label{Drudew.SF.RLL}
\end{table}

Fig. \ref{conductivity.SF.RLL} and Fig. \ref{conductivity.SF.attractive}
show the conductivity in the superfluid phase in the repulsive and 
attractive Luttinger liquid regions. In the Luttinger liquid the 
conductivity was predicted to increase with a power-law for small 
frequencies, and decay exponentially for big 
frequencies\cite{Giamarchi:1992,GiamarchiMillis:1992}. The conductivity
in the attractive region shown in Fig. \ref{conductivity.SF.attractive}
is in good qualitative agreement with this. In the repulsive case 
(Fig. \ref{conductivity.SF.RLL}) there are too few peaks to clearly
identify this behavior. Bigger systems would have to be studied to determine 
if the overall shape is qualitatively different from  the attractive region.

The Drude weight in the repulsive and attractive regime of the 
superfluid phase is shown in Tab. \ref{Drudew.SF.RLL} and 
Tab. \ref{Drudew.SF.ALL}. For some system sizes data with different 
numbers of states $m$ and broadening $\eta$ is shown. The numerical 
accuracy depends on these parameters, with bigger $m$ and smaller
$\eta$ for higher accuracy. The data in Tab. \ref{Drudew.SF.RLL} and 
Tab. \ref{Drudew.SF.ALL} shows that the impact of  $m$ and $\eta$ on
the Drude weight is small.
In both the attractive and repulsive case  we find big non-zero 
values that are close to the kinetic energy per site in the
systems. The differences in the Drude weight in different system 
sizes, with the exception of the smallest systems, are rather
due to numerical errors that grow with the system size, 
than due to finite-size effects.

\begin{table}[tb]
\caption{The Drude weight $D$ for different system sizes in the attractive
	region of the superfluid phase at density $\rho=3/4$, $t=0.5$,
	$U=1$,$V=0.4$. Same notation as in Tab.\ref{Drudew.MI}.
	}
\begin{tabular}{cccccccc}
	$L$ 	&$D/t$	&$\langle T \rangle / (t L)$
				&$2/t \int \sigma_1^{reg}(\omega) d\omega$
							&$\eta$	&$m$	&$\Delta$	&$\omega_{c}$\\
\hline
	32	&1.438	&1.444	&0.006	&0.2	&128	&$10^{-4}$&0.6\\
	64	&1.421	&1.427	&0.006	&0.2	&128	&$10^{-4}$&0.4\\
	128	&1.412	&1.417	&0.005	&0.2	&128	&$10^{-4}$&0.3\\
	128	&1.411	&1.417	&0.006	&0.2	&256	&$10^{-5}$&0.3\\
\end{tabular}
\label{Drudew.SF.ALL}
\end{table}

\section{Conclusions}
\label{Conclusions}

In summary, we have studied the phase diagram of the 
one-dimensional Bose-Hubbard model with on-site only interactions 
and with additional nearest-neighbor interaction. The
density matrix renormalization group (DMRG) was used to calculate 
chemical potentials for given densities and model
parameters, and by doing this for sets of parameters the
phase boundaries of the Mott-insulators and the charge density wave
phase were determined.

The low-energy behavior of the superfluid phase of one-dimensional
bosonic systems is that of a  Luttinger liquid. We determined the
Luttinger liquid parameter $K$ from the decay of the hopping 
correlation functions. Since the value of $K$ is known for
insulator-superfluid transitions, we could use it to locate
the Kosterlitz-Thouless transitions at the tips of the Mott-insulators 
and the charge density wave phase in the $\mu$-$t$ phase diagram. 

In the charge density wave phase we found that close to the phase transition 
the structure factor depends on a power-law of the superfluid 
correlation length. From this we conclude that there is a direct
phase transition from the charge density wave phase
to the supersolid, and  no intermediate
phase like a supersolid or normal phase.

The charge density wave phase is surrounded by a region of the
superfluid phase where $K>1$, which corresponds to a Luttinger liquid with
repulsive effective interactions.
Kane and Fisher have shown that this region will be turned into an insulator
by a single impurity. We determined the boundary of the repulsive
region by finding the line where $K=1$ in the phase diagram. We
found that this boundary does not go to $t=0$ as the Mott-insulator 
is approached, but ends in the side of the Mott-insulating region,
where the Luttinger liquid parameter also is $K=1$.

We calculated the ac-conductivity in the Mott-insulator and the 
superfluid phase. In the Mott-insulator and the attractive region
of the superfluid phase the ac-conductivity has the expected shape.
In the repulsive region of the Luttinger liquid we found a different
shape, but could not determine if this is due to the finite system
sizes. The Drude weight or superfluid stiffness was found to be
big in both the attractive and the repulsive region.

\section{Acknowledgments}

The authors would like to thank H.~Carruzzo, J.~K.~Freericks, T.~Giamarchi,
L.~I.~Glazman, V.~A. Kashurnikov, A.~J.~Millis and R.~T. Scalettar
for valuable discussions.
This work was supported by the National Science Foundation 
under grant DMR98-70930 and the DAAD
``Doktorandenstipendium im Rahmen des gemeinsamen 
Hochschulsonderprogramms III von Bund und L\"andern''.

\appendix

\section{Truncation of the model}

\begin{figure}[b]
  \begin{center}
    \epsfig{file={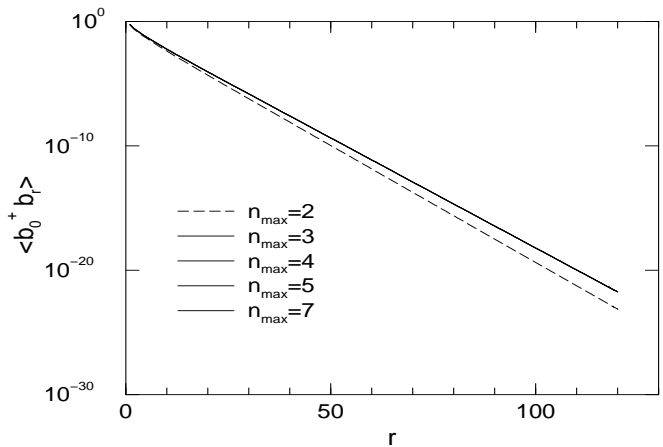}, 
  		height=\figheight,width=\figwidth, angle=-90}
    \caption 
    { 	
	The $\Gamma(r) =\langle b^\dagger_r b_0 \rangle$ correlation function
	for various truncations of the maximum number of particles per 
	site $n_{max}$ in the Mott-insulator at density $\rho=1$ in 
	a $L=128$ system with $t=0.1$  and $V = 0.4$. For $n_{max} \geq 3$ the
	different correlation functions become indistinguishable. 
    }
    \label{Corrfunc.with.trunc.MI}
  \end{center}
\end{figure}

The number of possible states per site in the Bose-Hubbard model
is infinite since there can be any number of particles on a site.
For practical DMRG calculations we truncate the model by only allowing
a maximum number of particles $n_{max}$ on each site. Pai et al. \cite{Pai:1996} chose $n_{max}=4$ in a DMRG study, while Kashurnikov and 
Svistunov\cite{Kashurnikov:1996} used $n_{max}=3$ in a Quantum Monte Carlo
study. To verify the effect of this truncation on the correlation function 
$\Gamma(r) = \langle b^\dagger_r b_0 \rangle$, we calculate systems
with different $n_{max}$. Fig. \ref{Corrfunc.with.trunc.MI}
shows $\Gamma(r)$ in the Mott-insulator. Due to the small particle-hole
excitations, the correlation functions are almost identical 
for $n_{max}\geq 3$.
In the superfluid phase there are more particle-hole excitations which are 
affected by the truncation. Fig. \ref{Corrfunc.with.trunc.SF} shows that
the correlation functions are independent of  $n_{max}$ for $n_{max}\geq 4$.
By choosing $n_{max} = 5$ for all calculations the effect of the truncation 
should be small enough not to affect the results presented in this work.

\begin{figure}[t]
  \begin{center}
    \epsfig{file={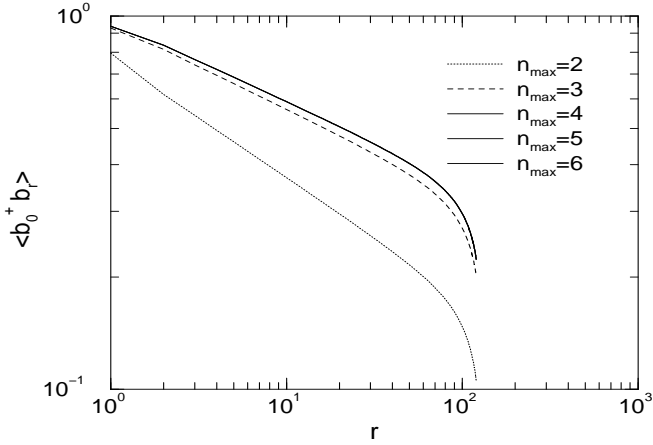}, height=\figheight,width=\figwidth, angle=-90}
    \caption 
    { 
	The $\Gamma(r) =\langle b^\dagger_r b_0 \rangle$ correlation 
	function for various $n_{max}$ in the superfluid phase in a $L=128$ 
	system with $\rho = 1$, $t=0.5$ and $V = 0.4$.
    }
    \label{Corrfunc.with.trunc.SF}
  \end{center}
\end{figure}

\section{Truncation of the DMRG basis}
\label{TruncationoftheDMRGbasis}

\begin{figure}[b]
  \begin{center}
    \epsfig{file={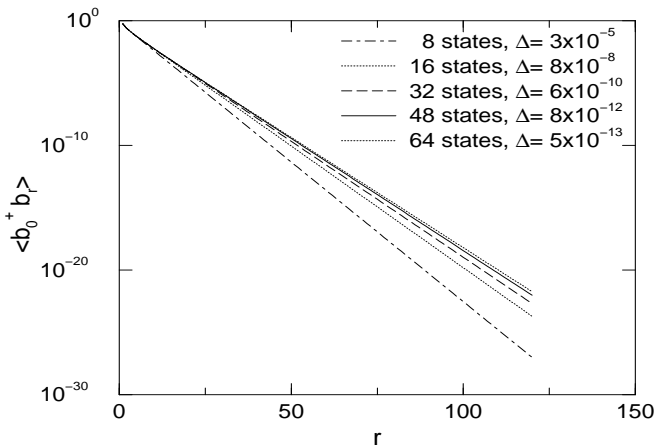}, height=\figheight,width=\figwidth, angle=-90}
    \caption 
    {
 	The $\Gamma(r) = \langle b^\dagger_r b_0 \rangle$ correlation
	function for various 
	truncation errors $\Delta$ in the $\rho=1$ Mott-insulator phase.
	System size $L=128$, $t=0.1$ and $V=0.4$.
    }
   \label{graph.bdagb.rho=1.t=0.1.various_discwt}
   \end{center}
\end{figure}

In every DMRG step the basis is truncated, and only the eigenstates of 
the density matrix with the biggest eigenvalues are kept. 
The density matrix weight of the discarded states $\Delta$ is a
measure of the error caused by these truncations.
To verify to which extent the truncation errors affect the
results, we calculate the correlation function $\Gamma(r) = \langle
b^\dagger_r b_0 \rangle$ with different numbers of states kept in the
DMRG basis. Fig. \ref{graph.bdagb.rho=1.t=0.1.various_discwt} shows
$\Gamma(r)$ with different truncation errors in the
Mott-insulator. Even for very small numbers of states the discarded
weight is very small, and the dependence on the weight of the
discarded states is weak. Note that the discrepancies are mostly
apparent due to the logarithmic scale.

\begin{figure}[t]
  \begin{center}
    \epsfig{file={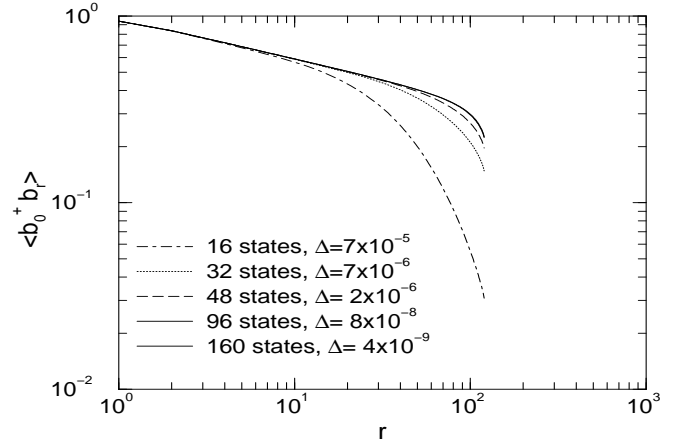}, 
		height=\figheight,width=\figwidth, angle=-90}
    \caption 
    { 	
	The $\Gamma(r) = \langle b^\dagger_r b_0 \rangle$ correlation 
	function in the superfluid phase at $\rho=1$ for various 
	truncation errors. System size $L=128$, $t=0.5$ and $V = 0.4$.
    }
  \label{graph.bdagb.rho=1.t=0.5.various_discwt}
  \end{center}
\end{figure}

\begin{figure}[b]
  \begin{center}
    \epsfig{file={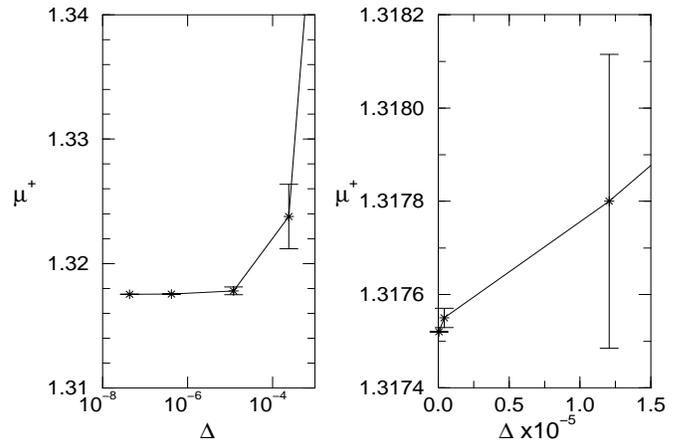}, height=\figheight,width=\figwidth, angle=-90}
    \caption 
    { 	$\mu^+$ versus the discarded weight $\Delta$ in the 
	Mott-insulator at $\rho=1$. The scale for $\Delta$ is logarithmic
	on the left plot and linear on the right plot.	
	The system size is $L=128$, $t=0.1$ and $V=0.4$. 
    }
	\label{graph.mup_and_mum_vs_trunc.t=0.1.rho=1}
  \end{center}
\end{figure}

The correlation function in the superfluid phase is shown in Fig. 
\ref{graph.bdagb.rho=1.t=0.5.various_discwt}. We find that the 
discarded weight with the same number of states is bigger than it 
is in the insulator. At small distances $r$ the correlation functions are very 
similar for all numbers of states, with increasing differences as $r$
is increased. If the discarded weights $\Delta$ are smaller than 
$8 \times 10^{-8}$, 
the correlation functions coincide even at the boundaries of the system. 
By requiring the discarded weight to be smaller than $\Delta \leq 10^{-9}$
for the calculation of the correlation functions, accuracy should be 
high enough in all cases.

The chemical potentials are calculated from the energies it
takes to add a particle or hole. Figures \ref{graph.mup_and_mum_vs_trunc.t=0.1.rho=1} and \ref{graph.mup_and_mum_vs_trunc.t=0.5.rho=1} show chemical
potentials calculated with different numbers of states kept versus
the discarded weight $\Delta$. The error bars correspond to the changes in
the energies during a DMRG sweep. Differences in the chemical potentials
are small for $\Delta<10^{-5}$.
We require the discarded weight to be smaller than $\Delta<5 \times 10^{-6}$
for the calculations of the chemical potentials, and to improve
the results further, we extrapolate linearly from the two values with 
the lowest $\Delta$.

\begin{figure}[tb]
  \begin{center}
    \epsfig{file={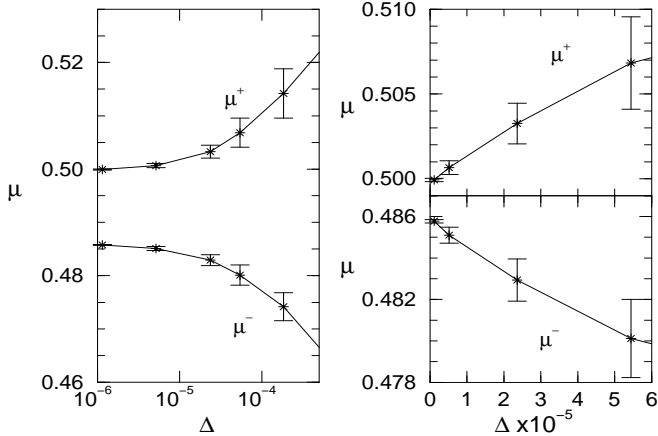}, height=\figheight,width=\figwidth, angle=-90}
    \caption 
    { 	$\mu^+$ and $\mu^-$ versus discarded weight $\Delta$
	 in the superfluid phase. The scale for $\Delta$ is logarithmic
	on the left plot and linear on the right plot.
	The density is $\rho=1$, system size $L=128$ and $t=0.5$ and
	$V=0.4$.
    }	
	\label{graph.mup_and_mum_vs_trunc.t=0.5.rho=1}
  \end{center}
\end{figure}

\section{Strong coupling expansion at the commensurate-incommensurate
	transition}
\label{StrongCouplingExpansion}

The fundamental difference between the commensurate-incommensurate 
phase transition at $\rho=1/2$ in one and two dimensions can
be illustrated with the help of a strong coupling expansion. 
In the strong coupling limit the kinetic energy is zero. The zero order
states are the groundstates of the Hamiltonian only including the 
particle-particle repulsion:
\begin{eqnarray}
\label{Hamiltonian.unperturbed}
  H  &=&  U \sum_{i} n_{i} ( n_{i} - 1 )/2
     + V \sum_{i}  n_{i} n_{i+1}\;.
\end{eqnarray}
The series expansion is made in terms of the kinetic energy term:
\begin{equation}
  H'  =  - t \sum_{\langle i,j \rangle} 
	( b^{\dagger}_{i} b^{\phantom{\dagger}}_{j} + h.c.)\;.
\end{equation}

Strong coupling expansions of this type have been successfully used to study
the phase diagram with on-site only 
interaction\cite{ElstnerMonien:1999,FreericksMonien:1994,FreericksMonien:1996}. 
To determine the phase boundaries of the Mott-insulator at $\rho=1$, first the
groundstate of Eq. (\ref{Hamiltonian.unperturbed}) has to be found. In  this
state there is simply one boson sitting on every site. Higher terms 
of the perturbation series introduce local particle-hole excitations. 
The chemical potentials on the boundaries can be determined
from the energy it costs to add a particle (Eq. (\ref{mucplus})) or a 
hole (Eq. (\ref{mucminus})). But in these cases, the zeroth order ground state
is degenerate, since the additional particle or hole can sit on any site.
This degeneracy is lifted in  first order perturbation theory. 
In first order the problem is reduced to the additional single particle 
moving on a uniform background of completely localized particles. 
Since the extra particle gains energy by hopping from site to site, it 
becomes completely delocalized. Although this behavior can be modified
in higher order of the perturbation series, and there are limitations 
due to the radius of convergence, it is interesting to note the 
difference between the perturbation series in the insulator and with an 
additional particle or hole. At
integer density the series starts out with a completely localized
state, while it starts with a completely delocalized state if there is 
an additional particle or hole. This is in good agreement with the fact that
there is a Mott-insulator for small $t$ at integer density, and a direct
phase transition to a superfluid (delocalized) phase if the density is changed.

A similar strong coupling expansion can also be used at the charge density 
wave phase at $\rho=1/2$. The charge density wave phase at $t=0$ 
is a state with alternating particle numbers, in one as well as two dimensions 
(Fig.\ref{pic.cdw.1d} and Fig.\ref{pic.cdw.2d}). Higher order terms in the
perturbation series introduce local particle hopping without 
destroying the charge density wave order.

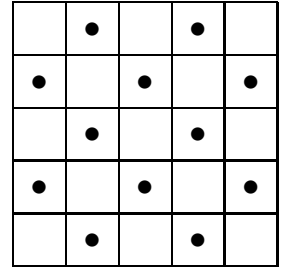
\begin{figure}[t]
  \begin{center}
   {
	\begin{picture}(100,100)(-1,-5)
	\matrixput(30,10)(40,0){2}(0,40){3}{\circle*{5}}
	\matrixput(10,30)(40,0){3}(0,40){2}{\circle*{5}}
	\matrixput(0,0)(20,0){6}(0,20){1}{\line(0,1){100}}
	\matrixput(0,0)(0,20){6}(0,20){1}{\line(1,0){100}}
	\end{picture}
     \caption{ The CDW in 2-d at $t=0$.}
     \label{pic.cdw.1d}	
   }
\end{center}			
\end{figure}
\begin{figure}[b]
  \begin{center}
   {
	\begin{picture}(200,20)(-1,-5)
\matrixput(6.5,6.5)(26,0){8}(0,13){1}{\circle*{5}}
\matrixput(0,0)(13,0){17}(0,1){1}{\line(0,1){13}}
\matrixput(0,0)(13,0){16}(0,13){2}{\line(1,0){13}}
	\end{picture}
    \caption{ The CDW in 1-d at $t=0$.}
    \label{pic.cdw.2d}	
   }
\end{center}			
\end{figure}
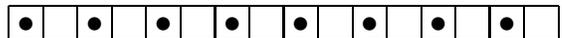

At the commensurate-incommensurate transition, an additional particle (or hole)
enters the system. For $V<U/2$ the energy is smallest if the additional 
particle goes to one of the empty sites. Fig. \ref{pic.cdw.p.2d} shows 
how the additional particle fits into the two-dimensional charge density wave. 
In higher orders of perturbation theory the additional particle, as well as
particle-hole excitations, hop on the charge-density background without
destroying it. From this we cannot infer if the true (non-perturbation
theory) groundstate is superfluid or not, and if the charge density
wave order is destroyed by the particle hopping. Nevertheless, it is 
interesting to note that for this case supersolids have been 
found\cite{Otterlo:1994,Batrouni:1995} in two dimensions. 
Close to the charge density wave phase at $\rho=1/2$, the charge density 
order survives at small doping. 

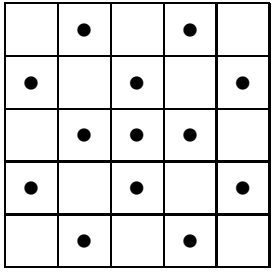
\begin{figure}[tb]
  \begin{center}
   {
	\begin{picture}(100,100)(-1,-5)
	\matrixput(30,10)(40,0){2}(0,40){3}{\circle*{5}}
	\matrixput(10,30)(40,0){3}(0,40){2}{\circle*{5}}
	\matrixput(0,0)(20,0){6}(0,20){1}{\line(0,1){100}}
	\matrixput(0,0)(0,20){6}(0,20){1}{\line(1,0){100}}
	\put(50,50){\circle*{5}}
	\end{picture}
     \caption{CDW with additional particle in 2-d at $t=0$.}
     \label{pic.cdw.p.2d}	
  }
  \end{center}			
\end{figure}

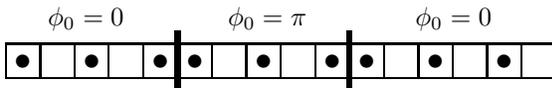
\begin{figure}[b]
  \begin{center}
   {
	\begin{picture}(200,30)(-1,-10)
	\matrixput(6.5,6.5)(26,0){3}(0,13){1}{\circle*{5}}
	\matrixput(71.5,6.5)(26,0){3}(0,13){1}{\circle*{5}}
	\matrixput(136.5,6.5)(26,0){3}(0,13){1}{\circle*{5}}
	\matrixput(0,0)(13,0){17}(0,1){1}{\line(0,1){13}}
	\matrixput(0,0)(13,0){16}(0,13){2}{\line(1,0){13}}
	\linethickness{2pt}
	\put(65,-5){\line(0,1){23}}
	\put(130,-5){\line(0,1){23}}
	\put(16,20){\text{$\phi_0=0$}}
	\put(84,20){\text{$\phi_0=\pi$}}
	\put(155,20){\text{$\phi_0=0$}}
	\end{picture}
    \caption{ Additional particle in 1-d at $t=0$. The thick lines are the 
 	domain walls. $\phi_0$ is the phase of the charge density wave.}
    \label{pic.cdw.addp.1d}	
   }
\end{center}			
\end{figure}

In contrast to this, the one-dimensional case looks quite different.
The additional particle also goes to an unoccupied site. If the charge
density wave remains unchanged, the additional energy is $\Delta E = 2 V$.
With the structure factor $S_\pi$, the charge density wave is given by 
an order parameter 
$\langle n_l \rangle = \rho + S_\pi \exp{(i \pi l + i \phi_0)}$.
An additional particle or hole can also be added by shifting the 
phase $\phi_0$ by $\pi$ over a region with an odd number of sites. 
Fig.\ref{pic.cdw.addp.1d} shows an example of such a state. 
In the center there is a domain with a $\pi$ phase shift, and the number of
particles compared to the charge density wave (Fig.\ref{pic.cdw.1d}) 
is increased by one. And the additional energy is also $\Delta E = 2 V$.
To lift the degeneracy between all these states in first order perturbation
theory, it can be seen that a particle hopping next to the domain boundary
is equivalent to the domain wall moving. Since energy is gained by this, 
the domain walls are completely  delocalized in first order perturbation
theory. Unlike the two dimensional case, where the charge density wave order
survives in all orders, in one dimension it is destroyed in first order.
While a perturbation series does not necessarily converge, this striking 
feature illustrates the fundamental difference between the one and two 
dimensional case.

\section{Current operator with open boundaries}
\label{Appendix:currentop}

\begin{figure}[b]
  \begin{center}
   \epsfig{file={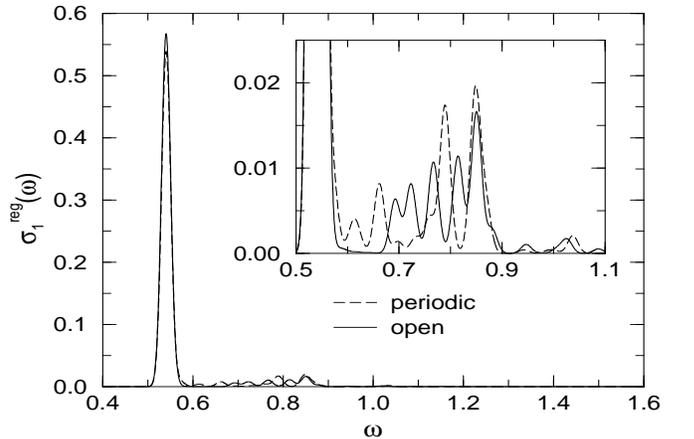}, height=\figheight,width=\figwidth, angle=-90}
    \caption
    {
	The conductivity $\sigma_1^{reg}(\omega)$ in the insulator
	with periodic and open boundary conditions. 
	The density is $\rho=1$, $t=0.05$, $U=1$ and $V=0.4$, system
	size $L=32$, and $m=128$ states.
	Broadening $\eta=0.2$ for the correction vectors,
	and $\eta_g=0.01$ for the plot. The inset shows the data
	on an expanded scale.
   }
  \label{Open_vs_Periodic.MI}
  \end{center}
\end{figure}

DMRG calculations work best with open boundary conditions. To calculate
the conductivity with DMRG, the current operator has to be implemented.
The current operator as it is given in Eq. (\ref{currentop}) can be
used directly with periodic boundary conditions, but to apply 
the current operator with open boundary conditions we modify it
with a filter function:
\begin{equation}
 j_{q=0} = \sum_{n=-\infty}^\infty P(x_n/M) (b^\dagger_{n+1} b_{n} - h.c. ) \;.
\end{equation}
The filter function $P(x_n/M)$ used here is a Parzen filter, $x_n$ is the
distance of site $n$ from the middle of the system, and $M=L/2$ is half the
system size. The Parzen filter looks very similar to a Gauss function, but
goes smoothly to zero at the system boundaries. It is given as:
\begin{equation}
\label{Parzen_filter}
        P(x) = a \left\{
        \begin{array}{lll}      
                1- 6 |x|^2 + 6 |x|^3 
                        & \text{if} & 0\leq |x| \leq 1/2 \\
                2 (1-|x|)^3 & \text{if} & 1/2 \leq |x| \leq 1 \; .
        \end{array}
        \right. 
\end{equation}

A prefactor $a$ is chosen to provide results with the same amplitude as 
those found in systems with periodic boundary conditions, with $a$ chosen
so that $\sum P(x_n/M)^2 = 1$. To verify the effect of open boundaries,
we do two separate calculations of the conductivity, one with periodic and
one with open boundaries, for otherwise identical system parameters. 
Fig. \ref{Open_vs_Periodic.MI} and 
Fig. \ref{Open_vs_Periodic.SF} show the conductivity in the Mott-insulator
and the superfluid phase with open and with periodic boundary conditions.
Even in the small systems with $L=32$ the curves are quite 
similar. In the superfluid phase a precursor peak at small frequencies can
be seen in the system with open boundary conditions. 
Fig. \ref{precursor.SF.RLL} shows how
the precursor peaks move to smaller frequencies as the system size is increased.
They are an artifact of the open boundary conditions, and we use frequency
cut-offs to exclude them from the calculation of the Drude weight. 
Tab. \ref{Drude.Comparisson.open_vs_periodic} shows the
Drude weights in the insulator and the superfluid. The values for
open and periodic boundary conditions compare
quite well, and we estimate an error of $\Delta D/ t \approx 0.02$.

\begin{table}[h]
\caption{The Drude weight $D$ with open and periodic boundary conditions.
	Systems size $L=32$, $t=0.05$, $U=1$ and $V=0.4$, $m=128$ states, 
	broadening $\eta=0.2$. Same notation as in Tab.\ref{Drudew.MI}.
	}
\begin{tabular}{ccccccc}
	boundary & $\rho$	&$D/t$	&$\langle T \rangle / (t L)$	
			&$2/t \int \sigma_1^{reg}(\omega) d\omega$
							&$\Delta$ & $\omega_c$\\
\hline
	periodic&1	&0.0195&0.6392&0.6197&$10^{-6}$	&-\\
	open	&1	&0.0017&0.6392&0.6375&$10^{-8}$	&-\\
	periodic&$3/4$	&0.8050&0.8957&0.0907&$10^{-4}$	&-\\
	open	&$3/4$	&0.7881&0.8836&0.0955&$10^{-6}$	&0.06\\
\end{tabular}
\label{Drude.Comparisson.open_vs_periodic}
\end{table}

\begin{figure}[tb]
  \begin{center}
   \epsfig{file={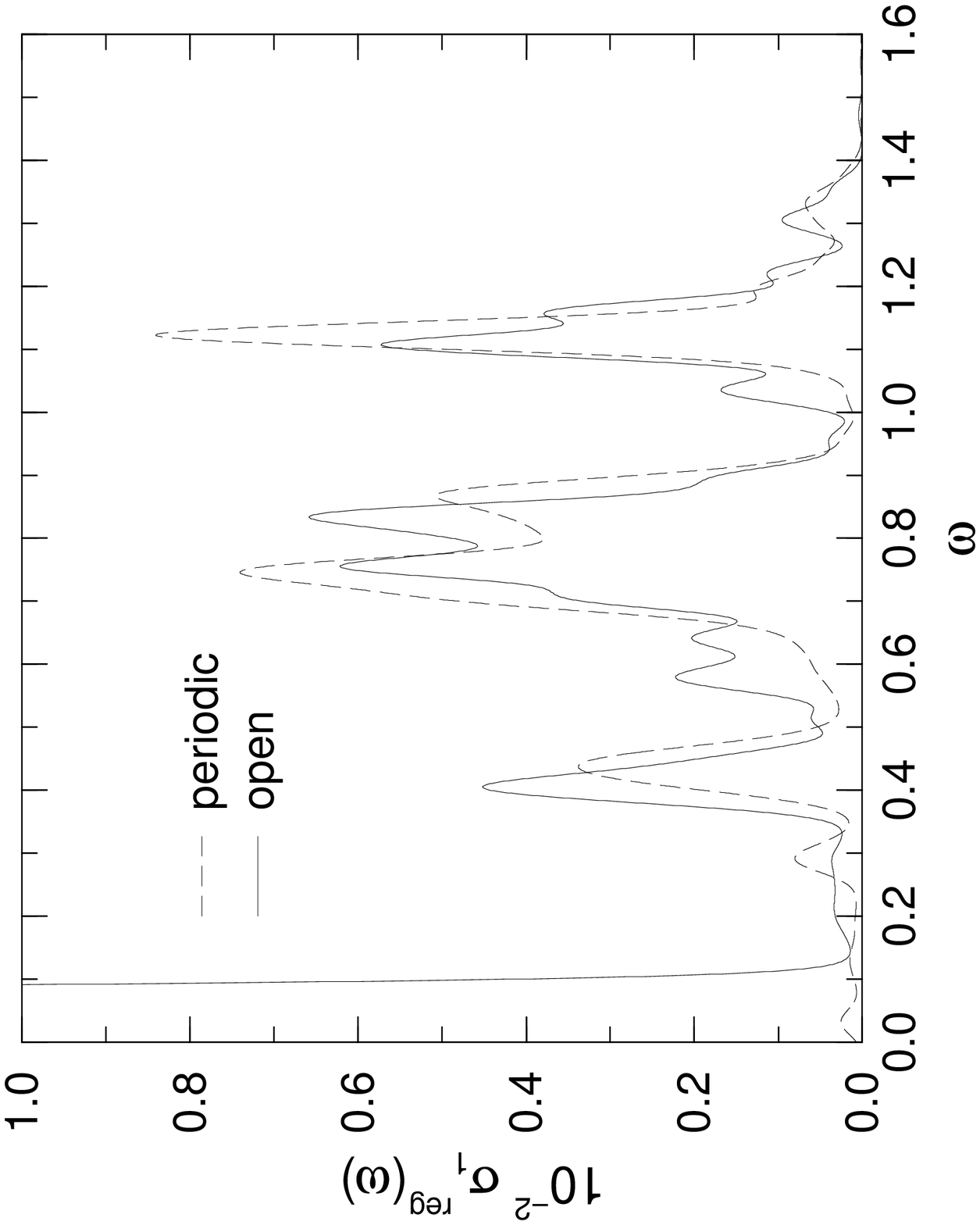}, height=\figheight,width=\figwidth, angle=-90}
    \caption
    {
	The conductivity $\sigma_1^{reg}(\omega)$ in the superfluid
	with periodic and open boundary conditions. 
	The density is $\rho=3/4$, $t=0.05$, $U=1$ and $V=0.4$, system
	size $L=32$, and $m=128$ states.
	Broadening $\eta=0.2$ for the correction vectors,
	and $\eta_g=0.02$ for the plot. The inset shows the data
	on an expanded scale.
   }
  \label{Open_vs_Periodic.SF}
  \end{center}
\end{figure}

\bibliographystyle{prsty}

\end{document}